\documentclass[useAMS,usenatbib,aas_macros]{mn2e}

\usepackage{epsfig}
\usepackage{lscape,graphicx,colortbl}
\usepackage{amssymb}
\usepackage{natbib}
\usepackage{times}
\usepackage{aas_macros}

\newcommand{\ifm}[1]{\relax\ifmmode#1\else$\mathsurround=0pt#1$\fi}
\newcommand{\kms}{\ifmmode\,{\rm km}\,{\rm s}^{-1}\else km$\,$s$^{-1}$\fi}

\newcommand{\hmsun}{\,\ifm{h^{-1}}{\rm M_{\odot}}}
\newcommand{\msun}{\rm M_{\odot}}

\newcommand{\dd}{{\rm d}}

\newcommand{\be}{\begin{equation}}
\newcommand{\ee}{\end{equation}}
\newcommand{\bea}{\begin{eqnarray}}
\newcommand{\eea}{\end{eqnarray}}
\newcommand{\z}{\emph{z}}
\newcommand{\mf}{M_{200}}
\newcommand{\fof}{{\scshape fof~}}
\newcommand{\mh}{M_{\rm infall}}
\newcommand{\zf}{\emph{z}_{\rm infall}}

\newcommand{\adf}{\alpha_{\rm df}}
\newcommand{\tdf}{t_{\rm df}}
\newcommand{\mha}{M_{\rm infall}^1}
\newcommand{\mhb}{M_{\rm infall}^2}
\newcommand{\mhs}{M_{\rm infall,i}^s}
\newcommand{\mhc}{M_{\rm infall,i}^c}
\newcommand{\ms}{m_{\star}}

\newcommand{\mfa}{M_{200}^1}
\newcommand{\mfb}{M_{200}^2}

\begin{document}

\title[The relation between haloes and galaxies]
	{Looking at the relation between haloes and galaxies under the lens}

\author[E. Neistein \& S. Khochfar]
{Eyal Neistein\thanks{E-mail:$\;$eyal@mpe.mpg.de} and Sadegh Khochfar
\\ \\
Max-Planck-Institute for Extraterrestrial Physics, Giessenbachstrasse 1, 85748 Garching, Germany
\\}


\date{}
\pagerange{\pageref{firstpage}--\pageref{lastpage}} \pubyear{2012}
\maketitle

\label{firstpage}


\begin{abstract}
We use an empirical approach to model the stellar mass of galaxies according to their host dark-matter 
haloes and subhaloes (`HASH' models), where each galaxy resides in a subhalo taken from a large $N$-body 
cosmological simulation. This approach allows us to study the mass relation between subhaloes and galaxies 
(MR) using various observational constraints at redshift zero: the weak lensing signal (WL), the two point 
auto-correlation function (CF), and the stellar mass function of galaxies (SMF). 
Our method is based on modeling the lensing signal directly from the cosmological $N$-body simulation, 
and should thus be more accurate (at least for massive objects) than other methods based on analytic halo models.
We find that the WL does not provide a strong constraint on the MR. The current observational
accuracy allows for more than a factor of 10 freedom in the subhalo mass of central galaxies, for a given stellar mass.
The freedom for satellite galaxies is much larger, providing a very poor constraint on the number-fraction of 
satellite galaxies (0.05 - 0.8). These results are not significantly modified when using both 
the SMF and WL as constraints. 
We show that for the most massive galaxies, observational constraints based on the CF with 0.1 dex errors, are 
equivalent to 0.05 dex error in the WL. For intermediate and low
mass galaxies the WL and CF constrain the MR in a different way. 
Although the WL is currently not adding much information at these masses,
it has the potential of being important using future, more accurate measurements.
The models found here do not match simultaneously the observed CF and WL signals, and show
a limited ability to match the WL \& SMF. We suspect that this is partially due to the cosmological model 
assumed here and we therefore adopt a mock WL signal through most of this work. 
In comparison to previous models in the literature, the method presented here is
probably more general, as it reveals a larger range of solutions for a given set of 
observational constraints. 
\end{abstract}


\begin{keywords}
galaxies: abundances; galaxies: formation; galaxies: haloes; galaxies: mass function; 
galaxies: statistics; cosmology: large-scale structure of Universe; gravitational lensing: weak
\end{keywords}


\section{Introduction}
\label{sec:intro}

The relation between haloes and galaxies marks the crossroad of two major pathways. On the one hand dark-matter
is accurately modeled by large $N$-body cosmological simulations, where the properties of haloes are quantified
in detail \citep[e.g.][]{Springel05,Klypin11}. Moreover, the large dynamical range in these simulations provides a full 
description of the dark-matter substructure that is believed to host galaxies \citep{Diemand07,Boylan09}. 
On the other hand, the properties of luminous galaxies can now be measured over large cosmological volumes, 
reducing the statistical uncertainties to a level of a few per cent \citep[e.g.][]{Drory09,Bernardi10}. 
Observational techniques for estimating physical properties of galaxies,
like stellar mass and star-formation rate, are becoming more and more reliable \citep[e.g.][]{Nordon12},
making it feasible to model the physical quantities of galaxies directly.

Various different methodologies have been developed in order to study the mass relation between galaxies 
and haloes (hereafter MR). Such models need to be flexible enough, so they can fit the observational
constraints to a high accuracy. They also need to be computationally efficient,
to allow quick exploration of the parameter space. Lastly, the model parameters should be directly 
related to the observational constraints (e.g. the stellar mass function and the auto-correlation function).
Due to the above, all the models used to study the MR in detail are based on empirical relations between 
haloes and galaxies.
However, in order to make reliable predictions on the MR, we need our models to be general enough,
so they can encompass the full ensemble of solutions. One of the main targets of this work is to emphasize
the importance of this issue.

Halo occupation distribution models (HOD) were the first to make strong predictions on the MR, and are being
widely used ever since \citep[e.g.][]{Jing98,Cooray02,Berlind02,Tinker05,Zehavi05,Zheng07,Tinker08,Zehavi11}.
These models assume that each halo hosts a number of satellite galaxies $N_g$, 
where $N_g$  follows a power law as a function of the host halo mass.
The galaxies are positioned at random locations within the halo, with a probability function that
follows a scaled version
of the dark-matter density profile. In order to predict observables like the abundance 
and clustering of galaxies, HOD models use the number density of haloes and their 
clustering properties by fitting results from large $N$-body simulations.

Since HOD models are based on analytical arguments, they are very useful for developing an insight on the model
ingredients, and how they affect the observed quantities \citep[e.g][]{Berlind02}. In addition, HOD models are relatively 
simple to understand, as they include just a few free parameters. These benefits also allow for detailed
studies on how the underlying cosmological parameters affect the model results \citep{Abazajian05,Yoo06,Tinker07,Baldauf10}, 
which are then used for making predictions for the cosmological parameters \citep[e.g.][]{Mandelbaum12}.

A variant of HOD, based on the conditional luminosity function (CLF), was developed by \citet{Yang03}.
Within this approach the properties of galaxies (e.g. the luminosity function) are matched 
for a given range of halo mass, using analytical fitting functions for the
contribution from central and satellite galaxies. Although this
parametrization is somewhat different than the usual HOD models, and includes different degrees of freedom,
the two approaches are very similar. CLF studies were used to interpret various types of observations 
\citep[e.g. mass-to-light ratios in clusters,][]{vdBosch03}, and to constrain the cosmological parameters
\citep{vdBosch12,More12,Cacciato12}.

Lastly, abundance matching models (ABM) populate galaxies at the positions of \emph{subhaloes} that are taken 
from large $N$-body cosmological simulations \citep{Vale04,Conroy06,Shankar06,Conroy09,Behroozi10,Moster10,Guo11,
Rodriguez12,Reddick12}. In standard ABM, the MR between galaxies and their host subhaloes is set 
by matching the luminosity 
function (or stellar mass function) of galaxies to the mass function of subhaloes. The only important detail lies
in the way the subhalo mass is defined. While the mass of central subhaloes is derived at the observed redshift, the mass
of satellite subhaloes is set by their mass just before becoming a satellite, which typically occurs a few Gyr before
the observed epoch.
ABM models have no free parameters, and seem to be too simplified at first sight. However, these models are usually
very successful in matching the mass function of galaxies (by construction), 
their auto-correlation function, and even weak gravitational
lensing measurements \citep{Tasitsiomi04}.

All the models above use some substantial assumptions that might restrict the nature of the accepted solutions.
Both HOD and CLF are based on specific parametrization. For example: the average number of galaxies within a given halo
mass ($N_g$), the functional shape of the conditional luminosity function, and the location of satellite
galaxies. Although most of these were tested against hydrodynamical and $N$-body simulations or group catalogs 
\citep{Kravtsov04,Zheng05,Mandelbaum05a,Simha12}, these simulations do not necessarily span all 
the possible physical scenarios of galaxy formation. 
This problem is much more severe in ABM models, as they assume a specific relation between satellite
and central galaxies, forcing both populations to follow the same MR \citep[see e.g.][]{Neistein11a}. 

The concern that some assumptions are too restrictive also arises when comparing different studies.
For example, both \citet{Yang12} and \citet{Moster12} have tried to self-consistently incorporate
the evolution of the MR with redshift. Although these
models assume a very different behaviour for satellite galaxies \citep[satellites are not allowed to form stars in]
[but grow significantly in Yang et al. 2012]{Moster12} both models are able to fit the data. 
This example indicates that more complex models do not necessarily capture all options, and some non-negligible
freedom might be found after changing the assumptions of each model above.

In our previous study \citep[][hereafter paper-$I$]{Neistein11b} we have proposed a 
synthesis between HOD and ABM termed `HASH' (an acronym for `halo and subhalo'). Within this approach 
we assign galaxies to subhaloes that are taken from a large $N$-body simulation. 
We allow the mass of satellite galaxies to depend
on both the halo and subhalo masses. In this way we use a more reliable estimate for the number and location of 
satellite galaxies than HOD, while allowing the halo mass to play a role in shaping the clustering of galaxies. 
In our implementation, we try to use the minimal set of assumptions possible: the mass relation for central
galaxies is not restricted to follow any functional shape, the dependence on both the halo and subhalo masses is
being constrained only slightly, the number of satellite subhaloes can change by using different dynamical friction 
estimates, and the location of satellite galaxies can be modified. All these ingredients 
aim at making our approach more general, and should allow us to find more possible models that fit the data.

In paper-$I$ we have shown that our approach points to a large degeneracy in the models that fit both the
stellar mass function of galaxies (SMF), and their projected two-point auto-correlation function (CF).
The natural next step would be to check how additional constraints of very different nature can restrict
the degeneracy of our models. In this work we add measurements of weak gravitational lensing (WL) as it
is claimed to provide strong constrains on models when combined with the CF \citep{Yoo06,Cacciato09,Leauthaud12,
More12,Cacciato12,Mandelbaum12}. Unlike previous studies, we will show below that using observational constraints
from WL and SMF can hardly limit the mass relation between haloes and galaxies. Using all constraints together
(i.e. WL, SMF and CF) is currently equivalent to using only the CF and SMF. However, we show that future WL observations
with increased accuracy at small stellar masses should be able to significantly contribute to our knowledge on the MR.

The general principle of the HASH approach is motivated by the properties of galaxies 
within a semi-analytic model (SAM). We have shown 
in \citet{Neistein11a} that the combination of the subhalo and halo mass can reproduce the clustering of the
SAM galaxies to a reasonable accuracy. In general, the subhalo mass is more closely related to the stellar mass, 
and gives a good handle on the location of galaxies, while the halo mass strongly affects the clustering properties 
of galaxies, and induce possible environmental effects \citep[e.g.][]{Khochfar08}.
Recently, \citet{Yang12} and \citet{Moster12} have pointed out the importance of using the infall 
redshift\footnote{The infall 
redshift is defined as the last time the satellite galaxy was the central object within its \fof group,
see Eq.~\ref{eq:minfall} below.}
in HOD and ABM models. In our approach, the stellar mass does not depend directly on the infall redshift.
However, for a given satellite subhalo mass, the host halo mass
correlates well with the infall redshift. Moreover, according to \citet{Neistein11a} the combination of both subhalo
and halo masses reproduces the clustering of SAM galaxies better than using the subhalo mass and infall redshift.

This work includes the following parts. In section \ref{sec:hash} we describe our methodology and
explain how we model the SMF, CF, and WL. In section \ref{sec:results}
we show the results for the mass relation between haloes and galaxies, and explore the contribution from
different constraints. Lastly, we summarize our main findings and discuss them in section \ref{sec:discuss}. 
Throughout the paper we write `Log' to designate `Log$_{10}$'.


\section{HASH models}
\label{sec:hash}

In this section we provide details regarding the method and models used in this paper. Our approach assigns a stellar
mass to a set of subhaloes from a large $N$-body simulation. The set of subhaloes, their mass and location are
described in the next three subsections. In the rest of the subsections we explain how the SMF, CF, and WL are derived,
and how we define specific models within our formalism. For a few example models that fit both the SMF and CF,
the reader is referred to paper-$I$.

\subsection{The cosmological simulation}

We use information on haloes and subhaloes from the Millennium simulation \citep{Springel05}. 
This simulation follows 2160$^3$ dark-matter particles (each with mass of $8.6\times10^8\hmsun$) 
within a box of length 500 $h^{-1}$Mpc. It is based on a cosmological model with parameters
$(\Omega_m,\Omega_\Lambda,\sigma_8,h)=(0.25,0.75,0.9,0.73)$, and includes 63 output snapshots spaced 
by $\approx 250$ Myr. The \fof algorithm \citep{Davis85} was used to identify haloes, and was then used as an input for 
the \textsc{subfind} algorithm \citep{Springel01} to identify subhaloes. 
The merger trees used here are those based on subhaloes, as described in \citet{Springel05}.

\subsection{Haloes and subhaloes}

The subhalo mass, $M_h$, is defined here to be the mass of all the particles inside a subhalo,
as identified by \textsc{subfind}. Throughout this paper we use the \emph{infall} mass of subhaloes,
$\mh$, which should correlate better with the stellar mass of galaxies,
\begin{equation}
\label{eq:minfall}
\mh = \left\{ \begin{array}{ll}
M_h & \;\;\;\textrm{if central within its \fof group}\\ \;\;\;\
& \;\;\;\;\;\;\;\;\;\;\;\;\;\;\;\;\;\;\;\;\;\;\;\;\;\;\;\;\;\;\;\;\;\;\;  \\
M_{h,p}(\zf) & \;\;\;\textrm{otherwise}
\end{array} \right.
\end{equation}
Here $\zf$ is the lowest redshift at which the main progenitor\footnote{
Main-progenitor histories are derived by following back in time the most massive
progenitor in each merger event.}
of the subhalo $M_h$ was the most massive within its \fof group, and $M_{h,p}$ is
the main progenitor mass at this redshift.

In addition to $\mh$ we will use the halo mass $\mf$, which is defined as the mass within the 
radius where the halo has an over-density of 200 times the critical density of the simulation. The halo 
mass for satellite galaxies within a group is defined to be the halo mass of the central object of 
that group. In general, $\mf$ includes mass not only from the central subhalo within a group,
but also from all (or some) of its satellite subhaloes. 

At each redshift we construct a catalog of subhaloes, including all subhaloes that are identified at this
redshift, and additional population of unresolved subhaloes (see below). The catalog includes the following
information: $\mh$ and $\mf$ (for fixing the stellar mass); the position (for computing the CF, \S\ref{sec:CF}); and the 
projected dark-matter density profile (for computing the WL signal, \S\ref{sec:lensing}). 
Unless stated otherwise,  we assume that all observations are fixed at $\z=0$, and compute the subhalo catalog
at $\z=0$.

\subsection{Satellite subhaloes}

We divide the population of subhaloes into three types,
\begin{itemize}
\item {\bf central subhaloes}: most massive subhaloes within their \fof group.
\item {\bf satellite subhaloes}: all subhaloes except central subhaloes.
\item {\bf unresolved subhaloes}: subhaloes that were last identified at higher redshift, and are added 
according to the dynamical friction formula (Eq.~\ref{eq:tdf} below). All the unresolved subhaloes are also satellite subhaloes.
\end{itemize}

Within the $N$-body simulation, satellite subhaloes lose their mass while falling into a bigger subhalo, 
and thus might fall below the resolution limit used by the \textsc{subfind} algorithm (20 particles). 
However, the galaxies that should reside inside these satellite subhaloes might live longer, as they are more dense, 
and thus less vulnerable to stripping. While constructing the merger trees, subhaloes that fall below the resolution
limit are considered to be merged into the central object. Consequently, this effect can modify the abundance of
subhaloes even at two orders of magnitude above the minimum subhalo mass resolved by the 
simulation \citep[e.g.][]{Neistein11a}.

In order to take this effect into account, we add to our subhalo catalog at $\z=0$ some subhaloes that 
have merged with larger subhaloes at a higher redshift. This is done by modeling
the time it takes a galaxy to fall into the central galaxy by dynamical friction. At the last time 
the satellite subhalo is resolved ($t_0$) we compute its distance 
from the central subhalo ($d_{\rm sat}$), and estimate the dynamical friction time using the 
Chandrasekhar formula,
\begin{equation}
\tdf = \adf \, \cdot \, \frac{1.17 V_v d^2_{\rm sat}}{G M_{h,2} \ln \left( 1+ M_{h,1}/M_{h,2} \right)} \, .
\label{eq:tdf}
\end{equation}
Here $M_{h,1}$ is the mass of the central subhalo, $V_v$ is its virial velocity, and $M_{h,2}$ is the mass of 
the satellite subhalo. Once a satellite subhalo falls together with its central subhalo into a larger group, we 
update $\tdf$ for both objects according to the new central subhalo. All subhaloes for which $\tdf+t_0$ is larger 
than the cosmic time at $\z=0$ are added to the catalog of subhaloes at $\z=0$. When using the total mass of 
the satellite galaxy instead of $M_{h,2}$, $\adf$ was found to equal $\sim2$ \citep{Colpi99,Boylan08,Jiang08,Mo10}. 
Since we only use the dark-matter subhalo mass here, and we would like to be careful when adopting limiting
assumptions within our models, we allow $\adf$ to vary between 0.1 and 10.

In order to model properly the location of unresolved subhaloes we use two options. First,
we use the location of the most bound particle of the last identified subhalo 
\citep[as was done by e.g.][]{Croton06}. This particle is followed over time even though its host subhalo does
not exist anymore. The particle location at $\z=0$ is taken directly from the $N$-body simulation. 
As a second option 
we derive an analytical model for the infall of a subhalo inside a central potential, 
\begin{equation}
d=d_{\rm sat} \left( 1 - \tau^p \right)^{1/q} \,.
\label{eq:subhalo_loc}
\end{equation}
Here $\tau=(t(\z=0)-t_0)/\tdf$ is the fraction of time spent out of all the estimated dynamical friction time 
until $\z=0$, 
$d$ is the distance we adopt at $\z=0$ from the central subhalo, and $p$, $q$ are constants.
The three dimensional location of the subhalo is using the assumption of radial infall.
For a full derivation of this model, the reader is referred to paper-$I$.
When searching for the models that fit observations we always 
try both options: the location of the most bound particle,
and the analytical model above.

\subsection{Modeling the stellar mass function}

In order to model the stellar mass function of galaxies we first compute the mass function of subhaloes
as found in the Millennium simulation and divided into central and satellite components:
\begin{equation}
\phi_c(\mh)=\frac{1}{V}\frac{\dd{N^c}}{\dd{\log\mh}} \,,
\end{equation}
\begin{equation}
\phi_s(\mh,\mf)=\frac{1}{V}\frac{\dd^2{N^s}}{\dd{\log\mh} \, \dd{\log\mf}} \,.
\end{equation}
Here $V$ is the volume of the simulation box, and $N^c,N^s$ are the numbers of central and satellite
subhaloes respectively.
For central subhaloes we need to compute the mass function as a function of $\mh$ only,
because there is no obvious physical reason to assume a dependence on both $\mh$ and $\mf$ (on 
average, the value of $\mf$ for central subhaloes is smaller by only $\sim0.08$ dex than $\mh$,  
with an RMS scatter of $\sim0.06$ dex). The mass function for satellite subhaloes is saved as 
a function of both $\mh$ and $\mf$, allowing us to model the stellar mass as a function of these two different
variables.

Next we would like to use the mass function of subhaloes to predict the stellar mass function of galaxies.
When comparing to the observed SMF, we need to compute the number of \emph{galaxies} that corresponds to a 
given range in $\ms$. Since the stellar mass is a function of the subhalo and halo masses,
$\ms=\ms(\mh,\mf)$, the number of galaxies within a given range in $\ms$ is the same as the number of
subhaloes within the corresponding area in the ($\mh,\mf$) plane. 
A simple way to parametrize regions within the ($\mh,\mf$) plane is by using the boundaries of each region.
We define $U_i^c$ to be the region boundaries for 
central subhaloes, and $U_i^s$ the boundaries for satellite subhaloes. 
To conclude, The number of galaxies within a given range of $\ms$ is simply the integral of $\phi$:
\begin{eqnarray}
\label{eq:N}
\lefteqn{
N = V\int_{U_i^c}^{U_{i+1}^c} \phi_c\, \dd{\log\mh} \, + } \\ \nonumber & &
V\int_{U_i^s}^{U_{i+1}^s} \phi_s \, \dd{\log\mh} \, \dd{\log\mf}   \,.
\end{eqnarray}
%

\subsection{Modeling the auto-correlation function}
\label{sec:CF}

The usual way to compute the CF
is to first populate a specific list of subhaloes from the simulation with galaxies, 
and only then to compute their CF. 
Since the CF is based on counting the number of pairs of subhaloes/galaxies, each such
computation is very demanding in terms of computer resources.
In our HASH approach we first compute the underlying number of pairs of 
subhaloes at all individual masses. Only at a later step we integrate the pair 
numbers in order to compute the CF
of galaxies. This method is extremely efficient when exploring many models, allowing us to compute the CF
for $\sim10^7$ different models.

We use the entire set of central subhaloes from the Millennium simulation and count the number of pairs into $\psi_{cc}$:
\begin{equation}
\psi_{cc}(\mha,\mhb, r ) = \frac{1}{V^2} \frac{\dd^3{N_p^{cc}}}{\dd{\log\mha}\, \dd{\log\mhb}\,\dd{\log{r}}} \,.
\label{eq:psi_cc}
\end{equation}
Here $\mha$, $\mhb$ are the infall mass of the first and second
subhaloes in the pair, and $r$ is the distance between these subhaloes within
the $x$-$y$ plane, taking into account the periodic boundary conditions (we use the $x$-$y$ plane
in order to compute the \emph{projected} auto-correlation function,
as is described below). In practice, we divide the range in Log$\mh$ and
Log $r$ into bins of 0.1 dex (except for $\mh<10^{12}\,\msun$ for which we use
bins of 0.02 dex), and save $\psi_{cc}$ as a multi-dimensional histogram. 

In a similar way we count the number of pairs for central-satellite
and satellite-satellite subhaloes,
\begin{equation}
\psi_{cs}(\mha,\mhb, \mfb, r ) \,.
\label{eq:psi_cs}
\end{equation}
\begin{equation}
\psi_{ss}(\mha,\mhb, \mfa, \mfb, r ) \,,
\end{equation}
Note that for satellite subhaloes the number of pairs
is saved as a function of both $\mh$ and $\mf$.
This is done in order to properly model the dependence of
stellar mass on $\mf$. 

The observed CFs used in this work are based on the auto-correlation function within
bins of stellar mass (hereafter 'domains'). We therefore need to compute the number of pairs $N_p$
for all galaxies within a range of stellar masses. Similarly to Eq.~\ref{eq:N} we need to integrate
the different $\psi$ components over the corresponding region within the $\mh-\mf$ plane:
\begin{eqnarray}
\label{eq:Np}
\lefteqn{
\frac{\dd N_p(r)}{\dd \log r} = V^2\int_{U_i^c}^{U_{i+1}^c}\int_{U_i^c}^{U_{i+1}^c} \psi_{cc}\, \dd^2{\log\mh} \, + } \\ \nonumber & &
V^2\int_{U_i^s}^{U_{i+1}^s} \int_{U_i^s}^{U_{i+1}^s} \psi_{ss} \, \dd^2{\log\mh} \, \dd^2{\log\mf}  \, + \\ && 
V^2\int_{U_i^s}^{U_{i+1}^s} \int_{U_i^c}^{U_{i+1}^c} \psi_{cs} \, \dd^2{\log\mh} \,  \dd{\log\mf} \,.
\end{eqnarray}

The projected two-point auto-correlation function,
$w_p(r)$, is then defined as the deviation
in the number of pairs from the average value per volume:
\begin{equation}
w_p(r) = \left[ \frac{L^2}{N^2} \frac{N_p(r)}{A(r)} - 1 \right] L \,.
\end{equation}
Here $A(r)$ is the 2-dimensional area covered by the radial bin $r$, $L$ is the size of
the simulation box in $h^{-1}$Mpc, and $N$ is the number of subhaloes (see Eq.~\ref{eq:N}).
The factor $N^2 A(r)/L^2$ corresponds to the number of pair galaxies with a separation $r$,
in case of a uniform random distribution of objects inside a 2-dimensional box of size $L$ (note
that we use the convention of double counting each pair when computing $N_p$, so $N^2$ corresponds
to twice the number of unique pairs).

\subsection{Modeling weak lensing}
\label{sec:lensing}

In order to compute the WL signal of our HASH models, we first compute the projected two-dimensional 
density profile for each \emph{subhalo} within the Millennium simulation.  This is done by projecting the 
entire database of particles into the X-Y plane, so the projected density profile of each subhalo includes
integration on the full simulation box along the Z-axis\footnote{We use a grid in the X-Y plane
with a cell size of 10 $h^{-1}$ Kpc.}. These profiles are then averaged out into bins of subhalo
and halo masses, resulting in:
\begin{equation}
\Sigma_c( \mh, r ) \,, \,\,\,\,\,   \Sigma_s( \mh, \mf, r ) \,.
\end{equation}
Here $r$ is the two-dimensional distance to the centre of the subhalo, and the subscripts s/c mark 
satellite and central subhaloes respectively. We save $\Sigma_{s,c}$ as a multi dimensional histogram, with
bins of 0.1 dex for each variable (except for $\mh<10^{12}$, where the bin size is 0.02 dex), as was done in 
the previous subsections.

The observed weak lensing signal is a transformation of the projected density profile:
\begin{eqnarray}
\lefteqn{\Delta\Sigma_c (\mh,r) \equiv} \\ \nonumber & & 
 \frac{2}{r^2}\int_0^r r_1 \Sigma_c(\mh,r_1) {\rm d} r_1 - \Sigma_c(\mh,r) \,,
\label{eq:delta_sigma} 
\end{eqnarray}
where the first term on the right hand side is the average profile within $r$.
Note that $\Delta\Sigma_c$ is a linear function of $\Sigma_c$, and 
is zero for $\Sigma_c$ that are constant in the variable $r$ \citep[the mass sheet degeneracy,][]{Gorenstein88}.
Consequently, integrating the density profile of each subhalo along the Z-axis of the entire simulation 
box does not affect the value of $\Delta\Sigma_c$. 
For satellite subhaloes, the derivation of $\Delta\Sigma_s$ uses the same identity as above, but with $\Sigma_s$. 
We refer to $\Delta\Sigma$ as the `lensing signal' or just 
'WL signal' throughout the paper.

As was done above (Eq.~\ref{eq:N}), in order to compute the model prediction for a range of $\ms$ we integrate
$\Delta\Sigma_c$ and $\Delta\Sigma_s$ over the values of $\mh$ and $\mf$ that correspond to each domain.
The integration uses the domain boundaries $U_i^c$ and $U_i^s$ exactly like in Eq.~\ref{eq:N}, with the only
difference in using $\Delta\Sigma$ instead of $\phi$.

\begin{figure}
\centerline{ \hbox{ \epsfig{file=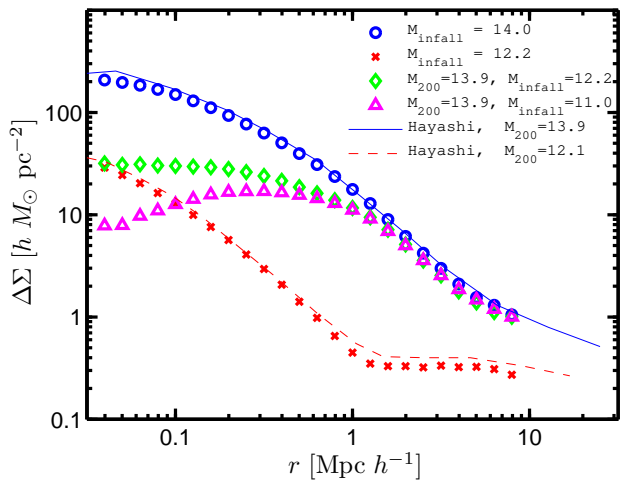,width=9cm} }}
\caption{The lensing profiles for different subhalo masses. \emph{Symbols} show
the average lensing profiles for central and satellite subhaloes within the Millennium simulation
with $\mh$ and $\mf$ as indicated (profiles of central subhaloes are those where the given mass is only $\mh$). 
For comparison we show in \emph{solid} and \emph{dashed} 
lines the same quantities for central subhaloes, taken from \citet[][note that for central subhaloes
$\mf$ is smaller than $\mh$ by $\sim$0.08 dex]{Hayashi08}. For satellite subhaloes we use here $\adf=3$,
and location of unresolved subhaloes following their most bound particle.
}
\label{fig:subhalo_profiles}
\end{figure}

In Appendix \ref{sec:app_prof} we list the specific values of $\Delta\Sigma_{s,c}$ for a large sample of 
profiles, allowing users to easily apply this data to their own use.
In Fig.~\ref{fig:subhalo_profiles} we show a few examples of profiles for both satellite and central subhaloes.
It can be seen that the profiles of satellite subhaloes are more flat than for centrals, but they coincide with
the profile of the central object at $r\gtrsim1$ Mpc. For central subhaloes, a factor of $\sim100$ in the subhalo
mass results in a factor of $\sim10$ enhancement in the lensing signal. Also, more massive subhaloes have more 
extended lensing profiles.

We also show in Fig.~\ref{fig:subhalo_profiles}
that the profiles taken from \citet{Hayashi08} agree well with ours. These authors have computed the same 
quantity as we do here based on the same simulation, but with a different technique, and only for central subhaloes. 
Note that for central subhaloes we use the $\mh$ mass, while \citet{Hayashi08} use the $\mf$ mass (on average $\mf$ 
is smaller than $\mh$ in $\sim0.08$ dex for the same central objects).

When constructing $\Sigma_s$ directly from the simulation
we only use the location of unresolved subhaloes following their most bound particle.
However, we have many models in which the location of these subhaloes is set by an analytic recipe 
(Eq.~\ref{eq:subhalo_loc}). In order to simplify the process of estimating
$\Sigma_s$ for such models, we do not compute the density profiles directly from the simulation 
(using the new location) but rather compute the density profile analytically, 
using the profile of the central object within the same group. 
This is done by computing the overlapping area between each radial shell of the satellite subhalo
and the shells of the central object. Each shell of the new
profile is therefore a combination of various shells from the central object, each with a different
weight factor, according to the size of the overlapping area.
As the central profile used
here is saved in radial bins, this method is accurate only for spherical density profiles, so it might introduce 
some errors in our case.

\begin{figure}
\centerline{ \hbox{ \epsfig{file=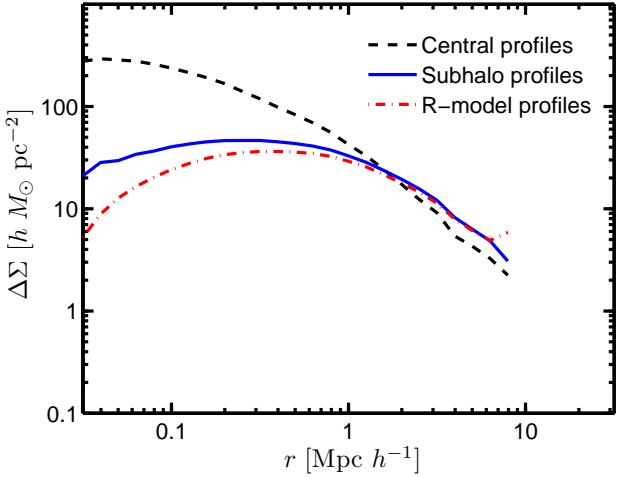,width=9cm} }}
\caption{The average lensing profile for unresolved subhaloes of mass $\log\mh$ within the range $[11,\, 11.5]$,
and $\log\mf$ between 14 and 15. Blue \emph{solid} line shows the actual profile
measured from the Millennium simulation, based on the location of the most-bound particle.
The red \emph{dotted-dashed} line shows the average profile of our estimated correction, 
in which each profile of a satellite subhalo is being computed from the central object profile, 
using the distance of the satellite. Black \emph{dashed} line shows the average profile of the central
object within the same \fof group for all these subhaloes. }
\label{fig:profile_offest}
\end{figure}

In order to test the method above, we compute analytically the profiles of unresolved subhaloes,
with the target location being the \emph{same} as is obtained by their most bound particle. We then check if the profiles
extracted from the simulation do agree with our analytic derivation.
In Fig.~\ref{fig:profile_offest} we show the results of this test. It is evident that this correction is not
entirely accurate, giving rise to an offset of up to a factor of 2 at the smallest scale. 
However, since the effect of unresolved subhaloes is rather weak on our results (see below),
the low accuracy shown here should not modify our results strongly.

The inconsistency of our approximation above means
that satellite subhaloes are located in regions of high density, also in comparison to the average location
within the host halo, at the \emph{same} radial distant from the centre. Consequently, when using a fully
analytical model, one needs to assume the locations of satellite subhaloes, as well as 
their excess density. In our approach, we neglect the excess density only for unresolved subhaloes (which are
a small fraction of all the satellite subhaloes), and only in cases where the location is not set by their
most bound particle.

\subsection{The definition of a model}

In this study we use five stellar mass bins (termed `domains') starting at $\log\msun=9.27$ and ending 
at 11.77, each with size of 0.5 dex. These domains were chosen because they match 
those of the observed CF used here. To simplify our formalism, we will use the same
domains to interpret the observations of WL and SMF.

As discussed above, in order to make predictions based on our models, we need to make assumptions
on how each domain in stellar mass is related to a region within the $(\mh,\mf)$ plane. Our assumptions
on these regions and their corresponding boundaries $U_i$ are as follows.

For central subhaloes (where we assume that $\ms$ depends only on $\mh$), 
each domain is defined as a range in $\mh$.  This means
that a range in stellar mass for central galaxies is equivalent to a range in $\mh$ for central subhaloes.
In other words, this parametrization simply assumes that the shape of the $\ms-\mh$ curve is determined by
a sample of a few points, allowing its shape to freely depend on the data. In order to define a model
we therefore need to fix one free constant for each boundary:
\begin{equation}
U_i^c =  \mhc \,.
\label{eq:alpha_i}
\end{equation}
Since we do not adopt prior limitations on $\mhc$ this description does not restrict the models in any sense.

For satellite subhaloes, we assume that each domain 
in stellar mass corresponds to a region of subhaloes that are located between
two boundaries within the $\mh-\mf$ plane:
\begin{equation}
\log U_i^s = \log\mhs + \delta_i \log\mf \,,
\label{eq:beta_i}
\end{equation}
where $\mhs$ and $\delta_i$ are free constants, 
$\mf$ is a variable and $U_i^s$ corresponds to the value of $\mh$
(in analogy to the equation of a straight line, $\,\!y=a+bx\!\,$).
This means that the boundaries of each domain are linear lines within the $\log\mh-\log\mf$ plane.
The values for $\mhs$ are restricted to follow our basic sampling bins. For $\delta_i$
we write $\delta_i = \tan \theta_i$ and sample $\theta_i$ in 
steps of 6.75 degrees, over the range [-90, 45]. 
Note that ABM models assume that $\delta_i$ are all zero, and $\mhs=\mhc$.

The parametrization we adopt for satellite subhaloes is using some prior assumptions
that might restrict all the possible functional dependencies of $\ms(\mh,\mf)$.
A more general model would allow each boundary to be a free contour within the $\mh-\mf$
plane.  Such models, however, are hard to handle as the number of possibilities for each domain would
be huge (in our simplistic case, and using the sampling density of our parameter space,
there are already $\sim10^{10}$ possible boundaries per each domain).
Nonetheless, our assumption on the functional shape of $U_i^s$ is only limiting the domain boundaries.
The combination of different domains, and the functional dependencies within each domain are free from any 
parametrization. To emphasize this point, \emph{any} function of the type:
\begin{equation}
\ms = f\left[ a(d)\log\mh+b(d)\log\mf \right] \,,
\end{equation}
is possible using our parametrization, with $a,b$ depending on the specific domain $d$ (i.e. $a$ \& $b$ can depend
slowly on $\mh$ as well), and $f(x)$ is completely free. This is different from ABM studies that assume 
a fix functional shape for $f$ and zero $b$.

Using domains in stellar mass, and corresponding regions in the $\mh-\mf$ plane saves
a large amount of computation time, since we perform our analysis separately on
each domain, reducing the number of free parameters by a factor of roughly five. 
During the post-processing phase the code accepts only models 
for which all the five domains have boundaries that coincide with their neighbour domains
(for example, the upper boundary of the first domain is the same as the lower boundary of the second domain).

Even though our parametrization introduces a large degree of 
freedom, our current approach does not include the effect of random scatter when assigning $\ms$ to subhaloes. Many works
have claimed that a log-normal scatter with $\sigma\sim 0.15-0.2$ is consistent with the observations. In our case,
we choose to deconvolve the scatter from the observational results, a method that is equivalent to adding a
scatter to the model \citep[see e.g.][]{Behroozi10}. We will use this method below, for checking the effect of 
scatter on the SMF and WL fits.

\begin{figure*}
\centerline{\psfig{file=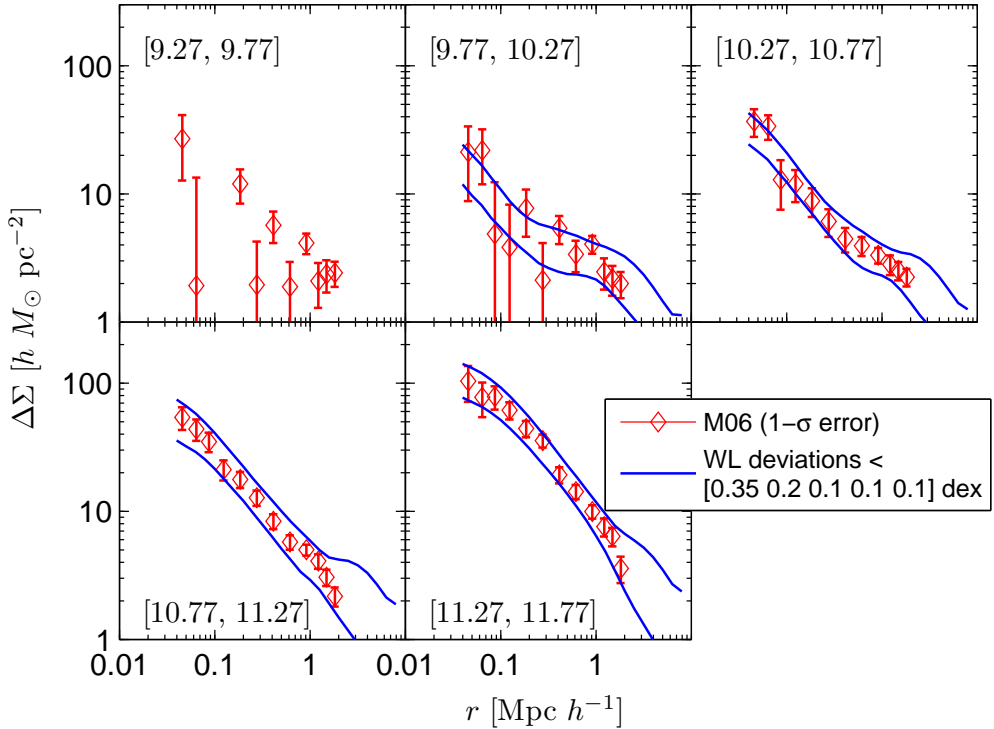,width=150mm,bbllx=30mm,bblly=80mm,bburx=188mm,bbury=200mm,clip=}}
\caption{The projected lensing signal derived for the models that fit WL only. Blue \emph{solid} lines 
represent the maximum and minimum of all the models that fit the observed WL values to a level of 
[0.35 0.2 0.1 0.1 0.1] dex RMS for each domain respectively.
The observational reference from M06 binned into our stellar mass domains are shown as \emph{symbols}, 
with error bars that correspond to 1 standard deviation. 
}
\label{fig:lens_prof}
\end{figure*}


\section{Results}
\label{sec:results}

\subsection{Search strategy and fitting criteria}

In this section we conduct a detailed search within all the possible HASH models, to see which model can
fit a given constraint. Our search goes over all the possible boundaries $U_i^c$ and $U_i^s$, taking into
account all the options for $\mhc,\,\mhs,\,\delta_i$ as defined 
in Eqs.~\ref{eq:alpha_i} and \ref{eq:beta_i}.
In addition, we have tested values of $\alpha_{\rm df}$ between 0.1 and 10, and a few options
for the location of unresolved subhaloes (using either the location of the most bound particle of each subhalo,
or various values of $p$ and $q$ from Eq.~\ref{eq:subhalo_loc}, as summarized in table A1 of paper-$I$).
In total we have 21 parameters for each model: 3 parameters that define 
each boundary times 6 different boundaries, 
and 3 general parameters that are assumed to be fixed over all domains: $\adf$, $p$ and $q$.
Overall, our search algorithm scans $\sim10^{11}$ models for each domain.

\begin{figure}
\centerline{ \hbox{ \epsfig{file=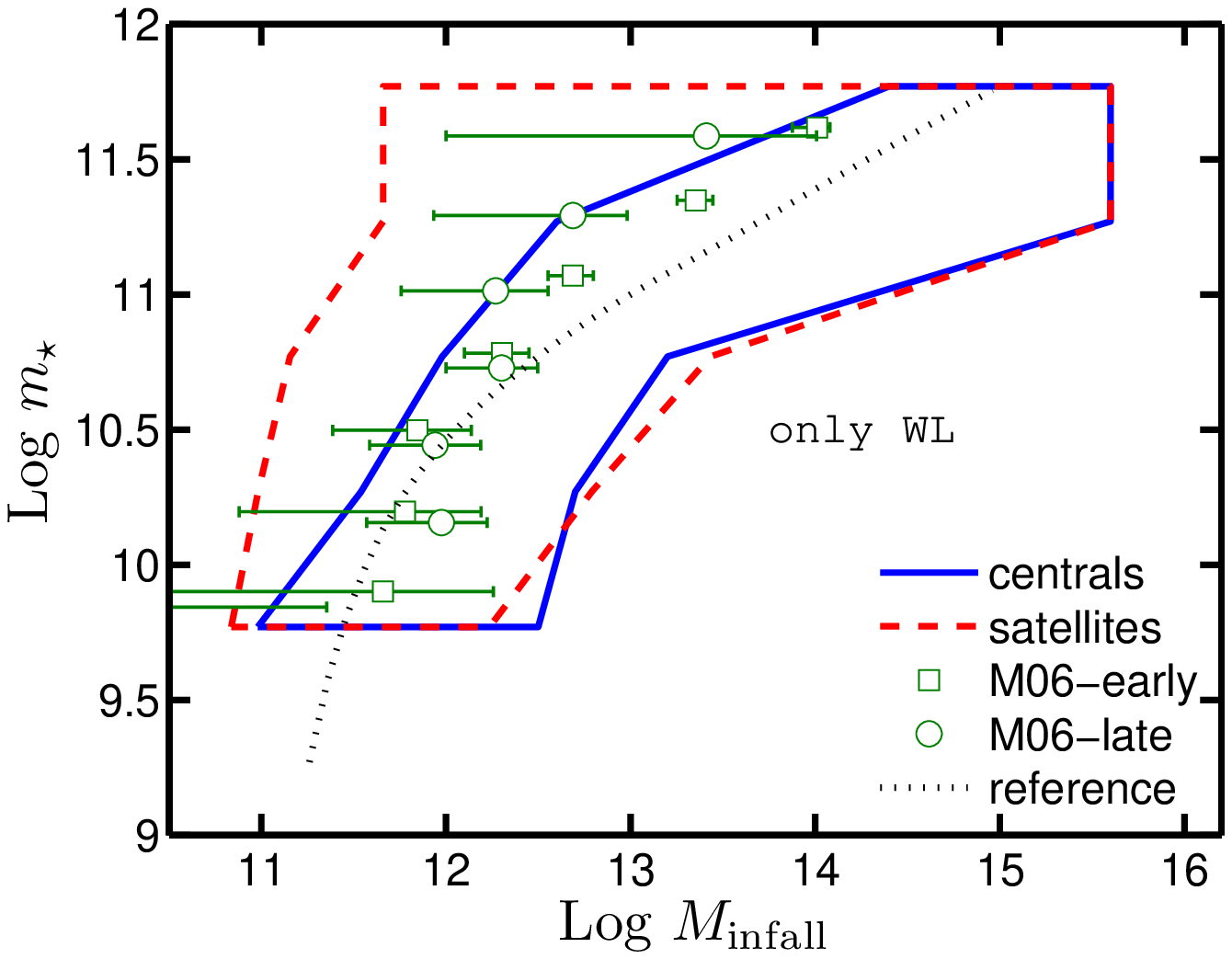,width=9cm} }}
\caption{The mass relation between haloes and galaxies using constraints from WL only. 
\emph{Thick} lines represent the maximum and minimum of all the models that fit the observed 
WL values to a level of [0.35 0.2 0.1 0.1 0.1] dex RMS for each domain. 
Results from M06 are plotted in \emph{symbols}, and are shown only for central galaxies. 
Squares and circles refer to the mean value of early and late type galaxies respectively, with error bars
that reflect 95 per cent confidence level. The fraction of late-type galaxies
out of the full sample is 0.74, 0.60, 0.46, 0.32, 0.20, 0.11, 0.05 (ordered in
increasing $m_\star$). For reference we plot in \emph{dotted} line the $\mh$ - $m_\star$ relation  
using both satellite and central galaxies as one population, with no dependence on $\mf$, and constrained
by matching the SMF from \citet{Li09}.
}
\label{fig:compare_mandel}
\end{figure}

When matching the SMF, we require an accuracy that is better than 20 per cent in reproducing the number 
of galaxies within each domain (Eq.~\ref{eq:N}). Once the model reproduces the total number of galaxies
within each domain (of size 0.5 dex in $\ms$) it is then always possible to distribute the galaxies properly
\emph{within} this domain to fully reproduce the shape of the SMF. This is doable because our parametrization
is not being limited within each domain. In addition, \citet{Leauthaud11} have shown that the SMF errors at different
masses are highly correlated, supporting the concept of fitting the SMF with only a few data points.

For the CF and WL we estimate the quality of the models 
using a simple RMS test:
\begin{equation}
{\rm RMS}_w = \sqrt{ \frac{1}{n}\sum\limits_{i=1}^{n} \left( \log w_{p,i} - \log \widetilde{w}_{p,i}  \right)^2 } \,,
\end{equation}
where $w_{p,i}$ is the observed CF value at the $i$-th point, $\widetilde{w}_{p,i}$ is the model prediction,
and the index $i$ goes over the points measured for one domain. A similar definition is used 
for estimating the quality of the WL signal. 

We note that when using $\log$ values of CF and WL, the observational
errors are roughly constant within each domain, making our RMS test similar to the popular $\chi^2$ test. 
The benefit in using an RMS test is that our results do not depend on the specific errors given
by the observational studies used here, allowing for an easy adaption for different data sets, and
simple comparison of errors in WL versus CF. In addition, the RMS criterion is not sensitive to the number 
of data points for each constraint, unlike the  $\chi^2$ test that includes all data points together.
Lastly, systematic
uncertainties in our methodology (e.g. the specific cosmological parameters used by the simulation; the level of accuracy
inherent in using two parameters to fix $\ms$) are not negligible, and it is not easy to estimate the full error bar
for a given data point. 

It is well known that various errors within the observed data sets are correlated. For example, \citet{Leauthaud11}
have studied these correlations in detail for the observations of SMF, CF, and WL. They show that CF and WL
values at different neighbouring scales are correlated, as well as SMF errors for neighbouring masses. In 
addition, correlations might exist between different observations. For example, the WL signal and CF will 
both be modified in a similar way by the value of $\Omega_m$ \citep{Yoo06}. Since we do not have the full covariance
matrix that describes correlations within all the data points, we choose to keep our fitting procedure simple
and transparent, and use the RMS criterion above.

To summarize, we list below all the criteria that are used to select models:
\begin{itemize}
\item The number density of galaxies within each domain agrees with the observed value at a level of 20 per cent.
\item The RMS estimate for the CF is computed by taking the observed points at $0.03 < r < 30 $ Mpc $h^{-1}$,
spaced by 0.2 dex in $\log\,r$. We use two different criteria, of RMS=0.1, 0.2 dex.
\item The RMS estimate for the WL is computed by taking all the observed points at $0.03 < r < 2 $ Mpc $h^{-1}$. 
Here we use two different criteria, an RMS level of [0.35 0.2 0.1 0.1 0.1] dex per each domain, or an RMS
value of 0.1 dex for all domains.
\item The number of galaxies more massive than the most massive domain (i.e. galaxies more massive 
than $10^{11.77}\,\msun$) can deviate by no more than 20 per cent, with respect to the nominal value plus/minus 
an additional poisson error. This larger error is adopted because there are usually only a few tens
of galaxies within this range. We note that even though galaxies at this mass range are not part of 
our five domains, we in practice fit the SMF at this mass range, similarly to what is done in ABM.
\item Models for which the fraction $\ms/\mh$ is bigger than the universal fraction of 0.17 are rejected.
\end{itemize}

\subsection{Models that fit weak lensing (WL)}
\label{sec:models_wl}

We first show how the WL can constrain the mass relation between haloes and galaxies when it is used as the only
constraint for the models.
The observations used for the search are the data from \citet[][hereafter M06]{Mandelbaum06}, 
re-binned to match the domains in stellar mass used for measuring the CF (these domains typically include
two mass domains from M06)\footnote{When re-binning the data from M06 we take into account the number 
of lenses within each original bin. We have also tested our search algorithm using the original bins from M06,
finding no significant difference.}. The observed data from M06
after re-binning can be seen in Fig.~\ref{fig:lens_prof}. Since the two lowest mass domains
are rather noisy, we allow their fit to deviate by up to 0.35 and 0.2 dex RMS respectively. For the other domains, 
we demand an RMS fit of 0.1 dex, which roughly agrees with an error of one standard deviation (the plotted error bars).

Technically, in addition to the WL constraint, we have required a factor of 50 accuracy in matching the SMF 
from \citet{Li09}. Unless we apply such a limit, the number of models gets too large for us to analyze. Nonetheless,
this constraint is rather weak, and should not affect the model results too strongly.

The results of our search in terms of the MR are plotted in Fig.~\ref{fig:compare_mandel},
where we plot the maximum and minimum $\mh$ for a given $m_\star$, using all the accepted models.
Note that for central subhaloes each HASH model predicts a unique MR, with no scatter. 
For satellite subhaloes, each HASH model might have a range of $\mh$ per a given $m_\star$, depending
on the mass of the host halo, $\mf$.
We therefore compute the median value of $\mh$ at each domain boundary $U_i$ ($m_\star$=9.27, 9.77, \dots, 11.77), 
and then plot in Fig.~\ref{fig:compare_mandel} the maximum and minimum for all these median values.
The set of models that are able to match the first domain ($9.27<\log(\ms/\msun)<9.77$) is too large 
for us to analyze, and is therefore not shown here.

Regarding central galaxies, our HASH models predict a significant level of freedom, of more than one order of 
magnitude uncertainty in stellar mass, for a given subhalo mass. Consequently, the WL signal is not useful
as a single constraint. We will show in the next section that this is also true when combining
constraints from WL and SMF. For satellite
galaxies, our search shows even larger uncertainties, probably because the profiles of satellite subhaloes have lower
values than for centrals, and less dynamical range (see Fig.~\ref{fig:subhalo_profiles}). Another hint for this is 
obtained from comparing the range of models above, against the range of models obtained where the locations of unresolved 
subhaloes are fixed at the position of their most bound particles. This reduction in the parameter space shows 
a small effect on the mass relation, hinting that unresolved subhaloes are not a key ingredient
in fixing the weak lensing signal.

\begin{figure*}
\centerline{\psfig{file=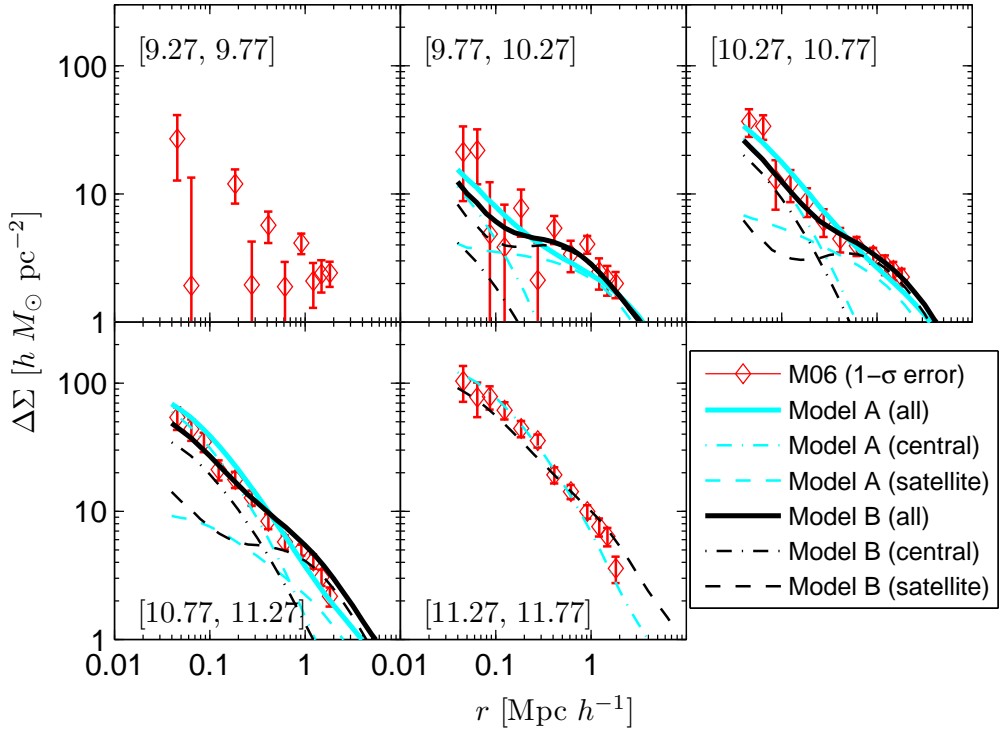,width=150mm,bbllx=30mm,bblly=80mm,bburx=188mm,bbury=200mm,clip=}}
\caption{The projected lensing signal for two example models that fit WL (as summarized in 
Figs.~\ref{fig:lens_prof} and \ref{fig:compare_mandel}).
\emph{Thick solid} lines are the total signals per each domain, \emph{dashed} and \emph{dotted-dashed} lines show
the contribution to the profiles from satellites and central subhaloes respectively. Each model is represented 
by a different line color. For the most massive domain, each model includes only one population of subhaloes 
(i.e. only central or only satellite subhaloes). For clarity, we do not show solid lines for this domain.
}
\label{fig:examples_wl}
\end{figure*}

Our results from Fig.~\ref{fig:compare_mandel} are consistent with the findings of M06 (shown 
for central galaxies). The fraction of massive late type galaxies (plotted in circles) is lower 
than $\sim10$ per cent, and has a negligible effect on the lensing signal. As a result, the large 
error bars for these galaxies do not point to a discrepancy in comparison to our results. It is 
interesting that our range of models is much larger than that 
of M06.  It seems that the methodology used here 
introduces more freedom in the models. Another option is that our
method of directly scanning the parameter space has some benefits over the Monte Carlo Markov chain approach 
used by M06. 
However, our fitting criteria are somewhat different from those adopted by M06, making it hard to compare
the two approaches directly.
Note that the cosmological model assumed here is similar to the one assumed in M06\footnote{M06 
have assumed the following set of parameters:
$(\Omega_m,\Omega_\Lambda,\sigma_8)=(0.3,0.7,0.9)$ in comparison to $(0.25,0.75,0.9)$ here.}.

In order to demonstrate the range of possible models that are able to match the WL, we show in Fig.~\ref{fig:examples_wl}
two different models that fit the WL as discussed above. As can be seen from this plot, different models might
have very different contributions from central and satellite galaxies. In the 2nd domain, model A is dominated
by centrals at small scales, while model B is dominated by satellites. At the most massive domain, the difference
between the models is maximal, and each model is governed by only one population of galaxies (central/satellite).

\subsection{Models that fit the SMF and WL}

\begin{figure}
\centerline{ \hbox{ \epsfig{file=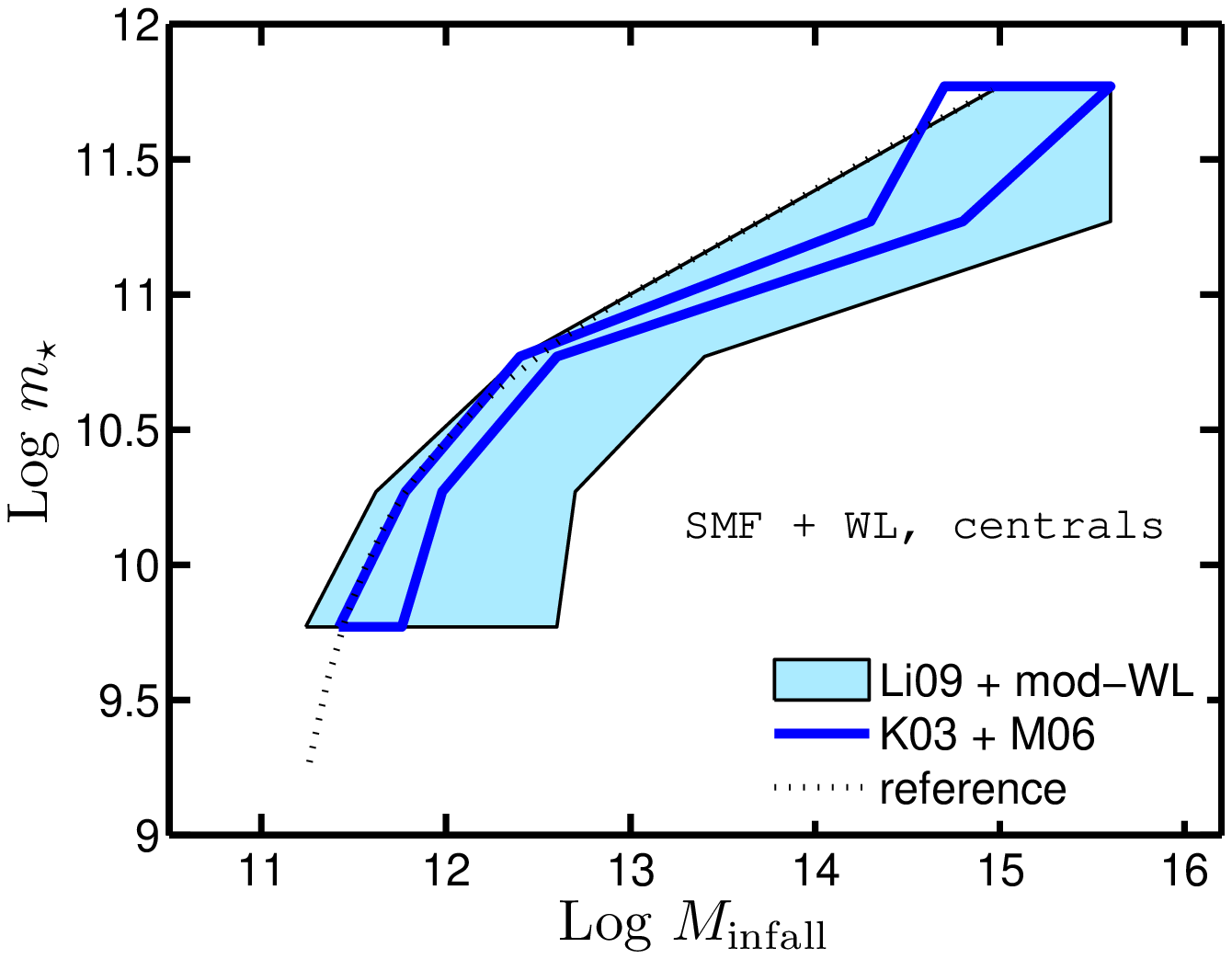,width=9cm} }}
\centerline{ \hbox{ \epsfig{file=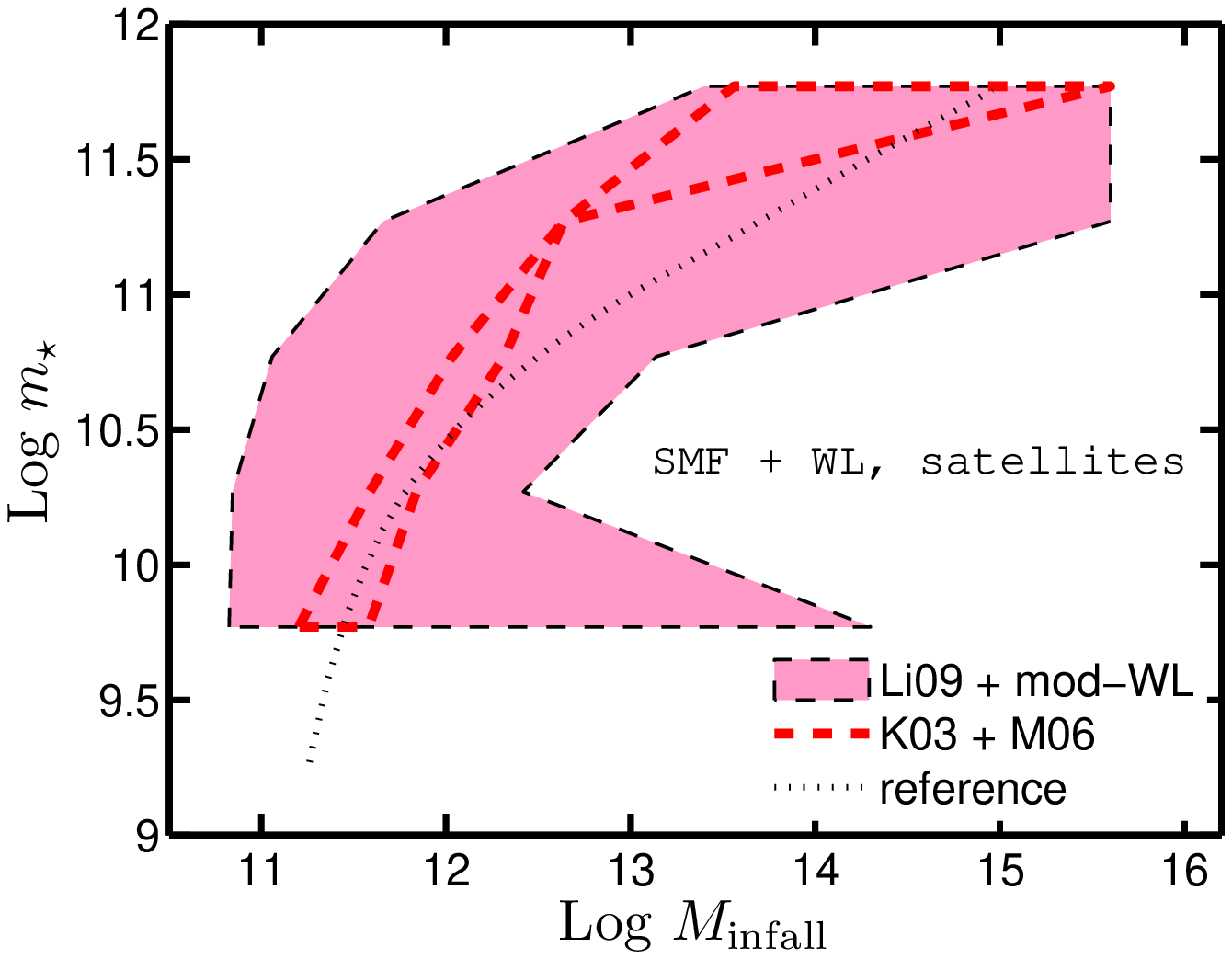,width=9cm} }}
\caption{The mass relation using constraints from both WL and SMF separated to central 
and satellite galaxies (\emph{upper} and \emph{lower} panels respectively). 
Fitting criteria for matching the WL signal are the same
as in Fig.~\ref{fig:compare_mandel}. The SMF is fitted to a level of 20 per cent for each domain. 
\emph{Lines} correspond to models that are constrained by WL from M06, and SMF based on
\citet{Kauffmann03}. The \emph{shaded} regions correspond to models that fit the 
SMF from Li09, together with modified WL signal (the modified WL signal is chosen to 
agree with Li09, see section \ref{sec:SMF_CF_WL}).  The \emph{dotted} line is plotted for 
reference, and is the same as in Fig.~\ref{fig:compare_mandel}. 
}
\label{fig:mass_relation_WL_SMF}
\end{figure}

The observed WL signal provides a relatively poor constraint on the MR.
It is therefore interesting to see if adding the SMF
would improve the results. Since we are going to use stellar masses from \citet[][hereafter Li09]{Li09} 
in the next section, we first apply their SMF constraint together with the WL of M06. 
Interestingly, even though the range of models that fit the WL is relatively large, 
we do not find \emph{any model} that matches the SMF constraint as well.

The difference in stellar masses between Li09 and \citet[][used by M06 for deriving the WL signal]{Kauffmann03}
was examined in the Appendix of Li09. It is shown there that the two mass estimates deviate by roughly 0.1 dex, where Li09
tend to assign lower masses for the same objects. This might contribute significantly to the discrepancy between
these two data sets found here. In order to bypass this issue we construct a
model search in which we fit the WL from M06 together with the SMF derived directly from
\citet[][we use the SMF version presented in Li09]{Kauffmann03}. The results
of this search are shown in Fig.~\ref{fig:mass_relation_WL_SMF}. The set of accepted models 
looks very narrow, and the models have high fraction of satellite galaxies at the massive end (for a given
value of $\ms$, the value of $\mh$ for satellite galaxies is much larger than for centrals). 

As will be discussed below, when using a mock WL signal, our results indicate to a much larger range of models, 
in comparison to the models obtained when using WL from M06. This issue raises the concern 
that our models suffer from a restrictive or wrong assumptions. We have made the following tests in 
order to study the possible effect of our assumptions:
\begin{itemize}
\item The mean redshift of lens
galaxies from M06 goes up with stellar mass, reaching $\z=0.19$ at the massive end. On the other hand,
our analysis assumes that the observations are all referring to $\z=0$. We have tested the importance of this 
effect by using lensing profiles of subhaloes taken at the proper redshift for each domain (while
using the same SMF from low redshift). Results for the MR in this case are similar to Fig.~\ref{fig:mass_relation_WL_SMF}.
However, a full exploration of this issue will require to allow the boundaries of neighbouring 
domains to differ, because the MR might evolve with redshift. Taking this effect into account deserves 
a further analysis that cannot be done here.
\item We have tested the range of models when using the original bins from M06, based on either the stellar mass,
or the luminosity \citep[luminosity based WL is combined with the luminosity function from][]{Blanton03}. 
In both cases the range of models is similar 
to what we plot in Fig.~\ref{fig:mass_relation_WL_SMF}.
\item We have tested the effect of adding a random scatter to the value of $\ms$. Assuming the scatter does
not depend on $\mh$, we have deconvolved the observed SMF and WL with a constant log normal scatter of 0.1, 0.15, 
and 0.2 dex. For each different scatter parameter, this results in a new SMF, and a new WL signal
(we deconvolve the WL signal as a function of $\ms$ for each radial bin, taking into account the
averaging of $\Delta\Sigma$ over the domain range in $\ms$). These data sets can then be fitted by our 
models with no scatter. In order to check the effect of this on the consistency between the SMF and WL,
we have computed new domains in $\ms$ that will yield the same number density of galaxies as in the 
original domains, but using the new deconvolved SMF. For these new domains we also generate the WL signal,
using the deconvolved signal above. 
Interestingly, even though both the SMF and WL change due to the addition
of scatter, the new domains have very similar WL profiles to the original domains, 
with deviations that are smaller than our fitting criteria.
Since our search procedure does not depend on the actual stellar mass value of 
each domain, it will produce the same solutions using the 
new domains above, as were found with the original domains.
The only effect of scatter would thus be to shift the domains of $\ms$ to lower 
values, especially at the high mass end. This will have no effect on the range of models presented 
in Fig.~\ref{fig:mass_relation_WL_SMF}.
\item The Millennium simulation used by our code assumes a higher value of $\sigma_8$ than the latest estimates 
\citep[0.9 instead of 0.8, see e.g.][]{Komatsu09}. The effect of various cosmological models was examined 
closely by a recent set of papers using the CLF approach \citep{vdBosch12,More12,Cacciato12}. From these
works \citep[see especially][]{More12} it seems that the combination of SMF and WL is less sensitive to the
cosmological model than the combination of SMF and CF. As we will show below, our model is able to match
both the CF and SMF to a good accuracy. This might indicate that the cosmological model assumed here is not
the reason for the bias we detect in the WL signal. However, since \citet{More12} have used a different model
than what we use here, this conclusion should be taken with a grain of salt. A clear test to this issue
can only be done by using a different $N$-body simulation with more accurate cosmological parameters.
\item We
define a `modified' WL signal, by adopting fiducial WL data points that are fully consistent with the 
observed CF \& SMF constraint used below. Although this WL signal is arbitrary, it provides a way to estimate
how the WL constraint will affect our results, in case it is fully consistent with the other observables.
We explain how we derive the modified WL signal in the next section.
In Fig.~\ref{fig:mass_relation_WL_SMF} we show the results of our model search using SMF based on Li09, together
with the modified WL signal. For consistency, we use the same error estimate for the WL, as was done above when
using the data from M06. This means that we demand an RMS accuracy of [0.35 0.2 0.1 0.1 0.1] dex for each domain.
Note that the \emph{range} of accepted models when using M06 together with \citet{Kauffmann03} is much more narrow
than the range when using the modified WL signal together with Li09. 
\end{itemize}
We conclude that the discrepancy between the SMF and WL (which is also valid when using the 
CF below) might be due to the limitations of our study. However, it might also be that
some observational systematics contribute to this effect.

From Fig.~\ref{fig:mass_relation_WL_SMF} it can be seen that using 
constraints from both the SMF and WL (the modified signal) is not so different from using only the WL
(as seen in Fig.~\ref{fig:compare_mandel}). The range in $\mh$ for a given stellar mass is still large for
central galaxies, and even larger for satellite galaxies. We conclude that using both the SMF and WL as constraints
on our models does not restrict the MR significantly.

\subsection{Models that fit the SMF, CF and WL}
\label{sec:SMF_CF_WL}

\begin{figure}
\centerline{ \hbox{ \epsfig{file=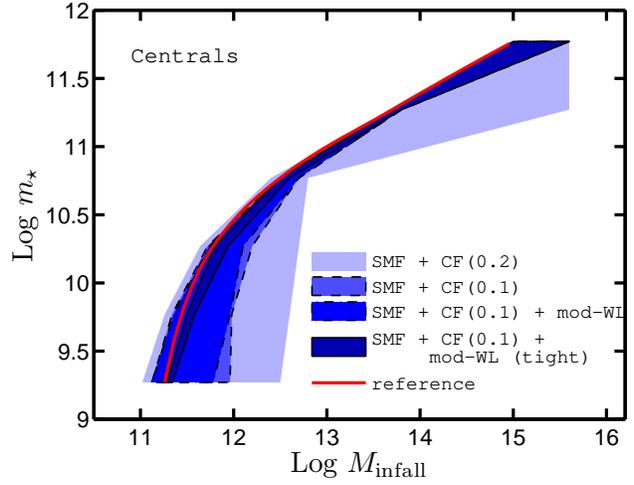,width=9cm} }}
\caption{The mass relation for central galaxies, using various sets of observational constraints. 
All models fit the SMF to a level of 20 per cent. The labels \emph{CF(0.1)} and \emph{CF(0.2)} refer 
to models that fit the CF to a level of 0.1 and 0.2 dex RMS respectively. The label \emph{mod-WL} corresponds
to models that match the modified WL signal to a level of [0.35 0.2 0.1 0.1 0.1] dex RMS, while \emph{mod-WL
(tight)} designate models that fit the modified WL signal to better than 0.1 dex RMS for all domains. 
All shaded regions correspond to the maximum and minimum $\mh$ values for each given $\ms$. The solid line
is given for reference and is the same as in Fig.~\ref{fig:compare_mandel}.}
\label{fig:models_summary_cent}
\end{figure}

\begin{figure}
\centerline{ \hbox{ \epsfig{file=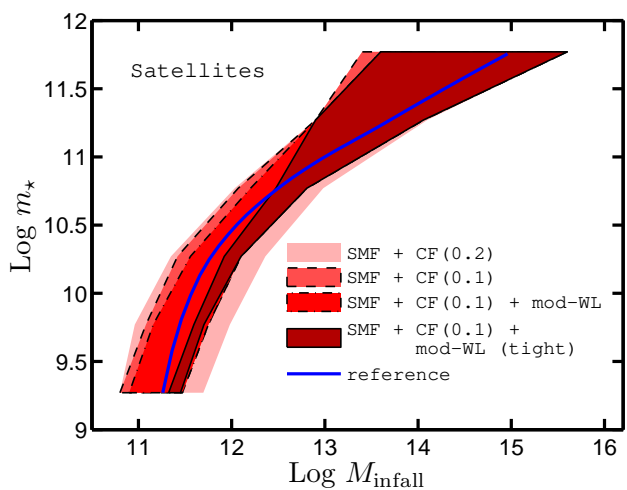,width=9cm} }}
\caption{Same as Fig.~\ref{fig:models_summary_cent}, but for satellite galaxies. Here we first compute the
median values of $\mh$ for each $\ms$ value. The shaded regions describe the maximum and minimum of the median
values for all models.}
\label{fig:models_summary_sat}
\end{figure}

\begin{figure*}
\centerline{\psfig{file=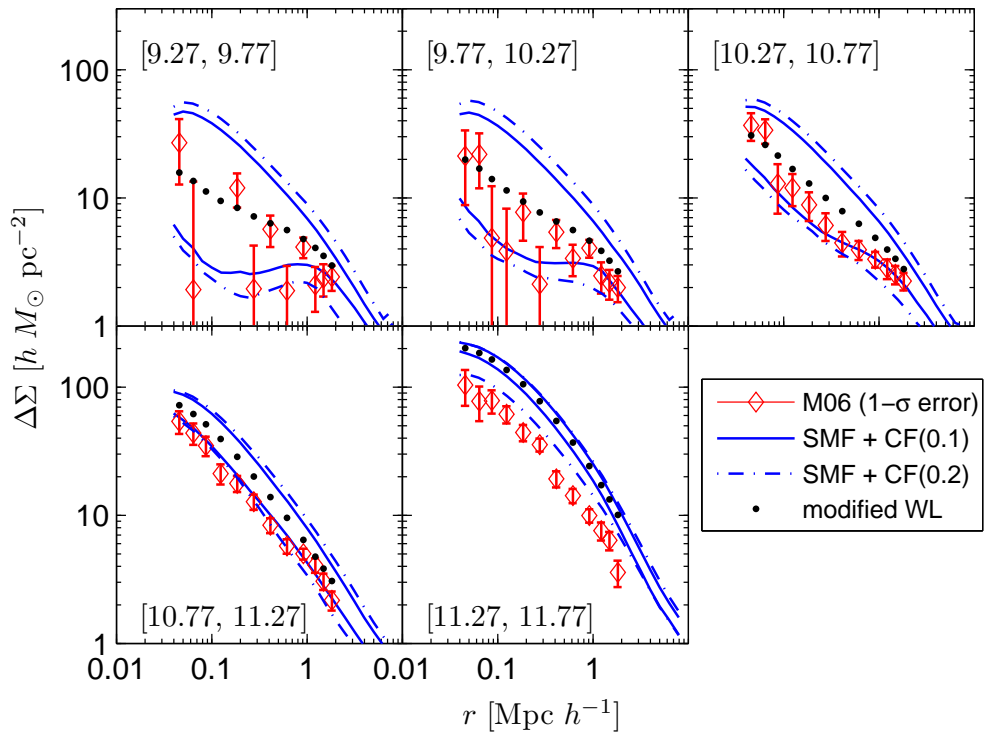,width=150mm,bbllx=30mm,bblly=80mm,bburx=188mm,bbury=200mm,clip=}}
\caption{The projected lensing signal derived for models that fit the CF \& SMF. Blue \emph{solid} lines represent the 
maximum and minimum of all the models that fit the SMF to a level of 20 per cent, 
and the CF at the level of 0.1 dex RMS.  Blue \emph{dashed-dotted} lines show how the range of models changes, when the CF
is constrained to fit the observations to a level of 0.2 dex RMS. The \emph{dots} show the modified WL signal we adopt
throughout the paper. These values are chosen to lie in the middle of the range of models that fit both the SMF and CF,
at radial locations that are the same as in M06. The observational reference from M06 are shown as symbols.
}
\label{fig:lens_prof_01}
\end{figure*}

In paper-$I$ we have made a search for all the HASH models that fit both the SMF and CF of galaxies.
The range of models and the fitting criteria tested there, are exactly the same as being used here.
The observed CFs are computed using the same stellar masses as in Li09, following the technique 
presented in \citet{Li06}. CF values are given within five domains in $\log(\ms/\msun)$: [9.27, 9.77],
[9.77 10.27], [10.27 10.77], [10.77 11.27], [11.27 11.77]. The CFs correspond to the projected 
two point auto-correlation function at 
scales ranging from 0.03 to 30 $h^{-1}$Mpc (the two most massive domains do not include data at
small scales).  Note that both the CF and WL signals go down to 0.03 $h^{-1}$ Mpc.

Computing the CF for each HASH model is much more time consuming then computing the WL signal, because
we need to integrate functions of higher dimensionality. As a result, we could not run a test in which only
the CF is used as a constraint, like was done here with the WL signal. There are two searches being done, 
each using both SMF and CF as constraints, but with fitting the CF to an accuracy of 0.1 and 0.2 dex RMS. 
In Figs.~\ref{fig:models_summary_cent}
and \ref{fig:models_summary_sat} we show the MR for these two different criteria. In comparison to using
the WL and SMF, we see that here the MR is more tight, especially for satellite galaxies. This however
might be due to either the smaller errors used to fit the CF (note that the observed CF values are based on SDSS DR7,
while M06 is based on DR4) or the larger range of scales probed by the CF. 

To summarize the results of fitting the CF and the SMF from paper-$I$, the range of models highly depends on the accuracy
by which we fit the CF. Models that fit the CF to a level of 0.2 dex RMS show a large range of MR, especially
for low mass central galaxies, reaching a factor of $\sim30$ uncertainty in $\mh$ for $\log(\ms/\msun)=9.27$.
This demonstrates that using the SMF as the only constraint does not yield tight MR within our formalism.
Demanding higher accuracy in matching the CF (to a level of 0.1 dex RMS) improves significantly the MR 
for central galaxies, but has a relatively minor effect on the MR for satellite galaxies. Although an accuracy
of 0.1 dex RMS is larger than the quoted observed error bars (these are usually between 0.03 and 0.06 dex,
depending on the domain and on the separation distance), we could not find models that match the CF to better 
than 0.08 dex.
This probably means that systematics errors contribute significantly to the fit quality. For more details on 
the models that fit the CF and SMF the reader is referred to paper-$I$.

\begin{figure}
\centerline{ \hbox{ \epsfig{file=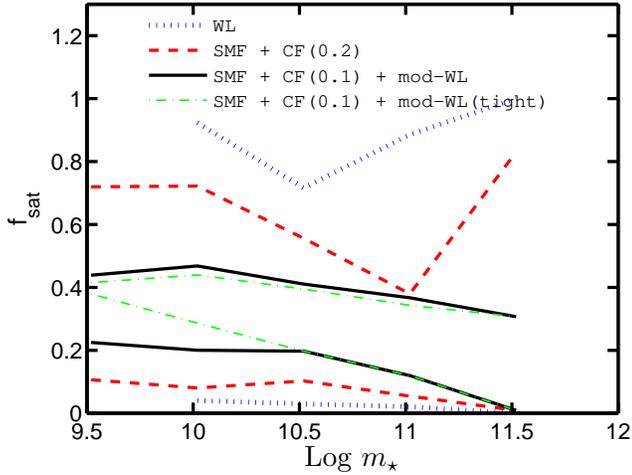,width=9cm} }}
\caption{The fraction of satellite galaxies out of all galaxies as a function of stellar mass.
The maximum and minimum values for all the models that fit each set of constraints are shown in
different line types as indicated. Labels are the same as in Fig.~\ref{fig:models_summary_cent}.}
\label{fig:sat_fraction}
\end{figure}

In order to understand better the interplay between the CF and WL constraints we compute the predicted WL
signal for all the models that fit the SMF and CF. The results of this test are shown in Fig.~\ref{fig:lens_prof_01}.
For the three low mass domains, the WL signal predicted by our models looks similar to the observed data from M06.
However, at the two highest mass domains, the models that fit the SMF \& CF are actually inconsistent with the observations
of WL from M06. Interestingly, the discrepancy still exist if we allow for a larger deviation from the observed CF,
at the level of 0.2 dex. All this agrees with the discrepancy between WL and the SMF discussed above. 
It should be noted that some part of the discrepancy seen in Fig.~\ref{fig:lens_prof_01}, especially at high stellar
masses, is due to the 
different stellar mass estimates used in Li09 in comparison to M06.

When combining constraints from SMF, CF and WL, the combination of the current data sets 
and our models is inconsistent. 
We have chosen to overcome this issue by modifying the WL data. Instead of using
the data from M06, we will now define a new mock WL data set, that agrees with the CF and SMF. At each domain
we define the `modified' WL signal by computing the average signal in $\log$ values, using the upper and lower 
limits of all the models that fit the SMF and the CF (we use a CF fit of 0.1 dex for this purpose). 
This modified WL signal is computed at the 
same radial bins as the original M06 data, and is plotted in Fig.~\ref{fig:lens_prof_01}.

In Figs.~\ref{fig:models_summary_cent} and \ref{fig:models_summary_sat} we show the MR when using constraints
from the SMF, CF, and WL. We use the SMF from Li09, CF as described above with an RMS fitting accuracy of 0.1 dex,
and the modified WL signal with fitting accuracy of [0.35 0.2 0.1 0.1 0.1] dex RMS. This combination reflects the
current accuracy of the various data sets. As can be seen from these figures, the range in MR is almost the same
as the range when using only the SMF and CF. We conclude that with the current uncertainties of the observed WL
signal, it adds very little to our knowledge of the mass relation between galaxies and subhaloes.

The error bars we assume for the WL signal are large, and are related to the fact that our reference data from M06
is based on DR4 from SDSS. In order to estimate the potential of future WL measurements we now demand a tighter fit to
the WL, at the level of 0.1 dex RMS for all domains. From Figs.~\ref{fig:models_summary_cent} and
\ref{fig:models_summary_sat} it seems that such an improved accuracy in the WL reference can significantly improve
the accuracy in the MR for low mass galaxies. However, the MR for massive satellite galaxies remains 
uncertain even in this extreme case, with an order of magnitude freedom in the value of $\mh$ for $\log(\ms/\msun)=11.77$
\citep[see the discussion in][related to WL]{vanUitert11}.

An accuracy of 0.1 dex is 10 times higher than the value of 0.35 dex we use with the current data. 
Assuming the errors in the WL signal are mostly statistical, this implies a factor of 100 increase in the 
number of galaxies in comparison to SDSS DR4 used here. Such a large number of galaxies is not expected to
be observed in the near future, but might be doable with future surveys (e.g. EUCLID).

Let us compare the power of the CF versus the WL in constraining the models. From the solid lines in 
Fig.~\ref{fig:lens_prof_01} (the most massive domain) it 
is evident that for galaxies with $\ms\gtrsim5\times10^{11}\,\msun$, an error of 0.1 dex in fitting the CF, 
corresponds to 0.05 dex range in the WL signal.
This means that for these massive galaxies, using a constraint of 0.1 dex for the CF (combined with 
the SMF) is equivalent to using a 0.05 dex error in the weak lensing signal. Consequently, in order for a new
WL measurement to be powerful, it needs to have much tighter error-bars than the existing CF measurements.

In Tables \ref{tab:summary_cent} and \ref{tab:summary_sat} we summarize the MR values obtained for the 
different constraints used in this work.

\subsection{Satellite fractions}

In Fig.~\ref{fig:sat_fraction} we show the fraction of satellite galaxies out of all galaxies in the five
different stellar mass domains used here. As was discussed above, models that are constrained by WL show
a very large range of fractions, spanning values between 0.05 and 0.8. This is a much larger range than what
has been obtained by previous studies (e.g. M06), again indicating that our formalism is
probably more general than
other existing models. When using all constraints (i.e. SMF, CF and the modified WL), the satellite fraction
are more constrained, reaching a range of $\sim$0.2 to $\sim$0.4. This range is  
larger than other estimates, especially at the high mass end \citep[see e.g.][]{Cacciato12}.

\begin{table*}
\caption{The possible range in the host subhalo mass $\mh$, for central galaxies of a given $\ms$, as derived from the 
various constraints used here (the same information is plotted in Figs.~\ref{fig:compare_mandel},
\ref{fig:mass_relation_WL_SMF} \& \ref{fig:models_summary_cent}). 
Column labels are the same as in Fig.~\ref{fig:models_summary_cent}.
All columns other than the first one correspond to the subhalo mass, $\mh$. All mass units are $\log\msun$}
\begin{center}
\begin{tabular}{lccccccc}
\hline $m_\star$ & WL$_{\rm M06}$ & SMF+WL$_{\rm mod}$ & SMF+CF$_{0.2}$ & SMF+CF$_{0.1}$ &  SMF+CF$_{0.1}$+WL$_{\rm mod}$ & SMF+CF$_{0.1}$+WL$_{\rm mod,tight}$ \\
\hline
9.27  & --            & --            & 11.02 - 12.50 & 11.12 - 11.96 &  11.12 - 11.78 & 11.28 - 11.34 \\
9.77  & 10.98 - 12.50 & 11.24 - 12.26 & 11.26 - 12.60 & 11.34 - 11.98 &  11.36 - 11.96 & 11.50 - 11.58 \\
10.27 & 11.54 - 12.70 & 11.62 - 12.70 & 11.64 - 12.70 & 11.72 - 12.20 & 11.74 - 12.10  & 11.80 - 11.94 \\
10.77 & 11.98 - 13.20 & 12.40 - 13.40 & 12.40 - 12.80 & 12.50 - 12.70 &  12.50 - 12.70 & 12.50 - 12.60 \\
11.27 & 12.60 - 15.60 & 13.70 - 15.60 & 13.70 - 15.60 & 13.70 - 13.80 & 13.70 - 13.80 & 13.70 - 13.80 \\
11.77 & 14.40 - 15.60 & 15.00 - 15.60 & 15.00 - 15.60 & 15.00 - 15.60 & 15.00 - 15.60 & 15.00 - 15.60 \\
\hline
\end{tabular}
\end{center}
\label{tab:summary_cent}
\end{table*}
\begin{table*}
\caption{Same as Table \ref{tab:summary_cent}, but for satellite galaxies. Note that here we show the
median $\mh$ value for each $\ms$, computed by going over all the possible values of $\mf$.}
\begin{center}
\begin{tabular}{lccccccc}
\hline $m_\star$ & WL$_{\rm M06}$ & SMF+WL$_{\rm mod}$ & SMF+CF$_{0.2}$ & SMF+CF$_{0.1}$ &  SMF+CF$_{0.1}$+WL$_{\rm mod}$ & SMF+CF$_{0.1}$+WL$_{\rm mod,tight}$ \\
\hline
9.27  & --            &  --           & 10.80 - 11.70 & 10.80 - 11.47 & 10.90 - 11.47 & 11.32 - 11.45 \\
9.77  & 10.84 - 12.24 & 10.83 - 14.30 & 10.96 - 11.98 & 11.08 - 11.75 & 11.16 - 11.75 & 11.59 - 11.70 \\
10.27 & 10.98 - 12.79 & 10.84 - 12.42 & 11.35 - 12.36 & 11.42 - 12.10 & 11.56 - 12.10 & 11.92 - 12.10 \\
10.77 & 11.15 - 13.43 & 11.06 - 13.14 & 12.03 - 12.98 & 12.07 - 12.81 & 12.20 - 12.81 & 12.47 - 12.81 \\
11.27 & 11.66 - 15.60 & 11.66 - 15.60 & 12.90 - 14.10 & 12.90 - 14.06 & 12.90 - 14.06 & 12.90 - 14.06 \\
11.77 & 11.66 - 15.60 & 13.40 - 15.60 & 13.40 - 15.60 & 13.40 - 15.60 & 13.60 - 15.60 & 13.60 - 15.60 \\
\hline
\end{tabular}
\end{center}
\label{tab:summary_sat}
\end{table*}
%

\section{Summary and discussion}
\label{sec:discuss}

In this work we have developed a formalism (termed `HASH') to interpret various observations regarding
the mass relation between haloes and galaxies (the `MR'). Our approach is using 
a set of subhaloes from a large cosmological $N$-body simulation, assigning one galaxy to each subhalo.
The stellar mass of galaxies is assumed to depend on both the subhalo and host 
halo masses ($\mh$ and $\mf$ respectively), allowing us to treat differently satellite and central galaxies.
We devote a specific attention to satellite subhaloes that are stripped within the $N$-body
simulation, but might still correspond to observed galaxies. We allow some freedom in the number
of these subhaloes via a model of dynamical friction, with a free constant scaling factor. The location
of these unresolved subhaloes has a non-negligible degree of freedom as well.

Our approach tries to adopt as much information from the simulation as possible, while allowing
freedom in quantities that are not accurately modeled. We thus hardly assume any prior on
the \emph{shape} of the MR. The only limitation built into our formalism assumes that the stellar mass of satellite
subhaloes is a function of the linear combination of $\log\mh$ and $\log\mf$, and this only at 
six different values of stellar mass.
Therefore, the MR for satellite and central galaxies vary significantly for different models within
our formalism. In addition, we are able to scan systematically the full parameter space 
to a high resolution, consistent with our fitting criteria ($\sim10^{12}$ models),
avoiding the uncertainties related to complicated search algorithms within the parameter space.

The observational constraints used here are: the stellar mass function \citep[SMF, we use both][]{Kauffmann03,Li09};
the two-point auto correlation function \citep[CF, we use a new version of][based on SDSS DR7]{Li06}; and a measurement
of galaxy-galaxy weak gravitational lensing \citep[WL, based on SDSS DR4 from][]{Mandelbaum06}. 

We claim that our formalism allows for more freedom in fitting the observed data sets than 
other methods. We explicitly show that when matching the WL signal, our method predicts a huge
amount of freedom in the MR, reaching a factor of 100 uncertainty in $\mh$ for a stellar mass of 
$\sim10^{11}\msun$ (see Fig.~\ref{fig:compare_mandel}) and number-fraction of satellite 
galaxies that lie between 0.05 and $\sim0.8$ for all stellar masses (Fig.~\ref{fig:sat_fraction}).
These result are very different than what was claimed in the past based on different approaches 
\citep{Guzik02,Hoekstra04,Mandelbaum06,Hoekstra07,vanUitert11}, although fitting criteria somewhat
differ between different studies. Another example is that our models are 
hardly being constrained by the SMF, in contrast to
halo occupation distribution (HOD) models for which the SMF plays a key role
\citep[e.g.][]{Leauthaud11}. 

Since all models used in this field are purely empirical, 
most of the assumptions being made are motivated by either more complex models, like 
hydrodynamical simulations, or by observational trends.
Nonetheless, different assumptions might highly affect the generality of the model, and 
might induce significant limitations to the set of solutions found to match the data. Therefore, it might be that
extending our formalism would further increase the set of models that can fit the data.  
On the other hand, other existing models might suffer from a limiting 
set of assumptions as well. For example,
HOD models assume that the number of satellite galaxies within a halo scales as a power law in the halo
mass. This might not be related directly to the physical processes of galaxy formation, and might not be valid in 
\emph{all} the possible scenarios of galaxy formation. We argue that a much careful
treatment of the assumptions made by the models should be considered, in order to obtain the most general formalism
possible.

The HASH formalism used here is specifically different from most previous studies in the way the WL 
signal is being modeled \citep[for more similar works see][]{Tasitsiomi04,LiR09}.
Here we compute the density profile within the simulation around each subhalo. We then stack all profiles 
of the same subhalo and halo masses $(\mh,\mf)$. This in principle provides a more accurate model than the 
usual, analytical HOD approach, at least for massive haloes that are well resolved within our simulation. 
According to our analysis, the observed WL signal from \citet{Mandelbaum06} is 
not fully consistent with respect to the SMF based on the same stellar masses.
We suspect that this is due to either the specific cosmological model assumed here, or 
to the non-uniform redshift of the data that cannot be modeled easily.
Consequently, when combining constraints from SMF, CF, and WL, 
we use a `modified', mock WL signal, that agrees with both the SMF and CF used here.

In \citet{Neistein11b} we have applied the constraints of both the SMF and CF, and showed that a 
large range of HASH models can match the data. The degeneracy
in those models corresponds to $\sim0.8$ dex freedom in the MR for low mass galaxies and for satellite galaxies of 
all masses. When we apply the additional WL constraint here, this range of models is hardly modified
(see Tables \ref{tab:summary_cent} and \ref{tab:summary_sat} for exact numbers). 
Based on the modified WL signal we claim that future WL observations with an uncertainty of $\sim0.1$ dex, would
be powerful in constraining the mass relation of low mass galaxies. 
Future surveys like EUCLID are therefore crucial in reaching such high accuracies.
For massive galaxies, it seems that constraints
based on the CF are more powerful than those based on WL, for the same error estimates.

There are various caveats within our approach. First, we use an $N$-body simulation based on a non-accurate 
set of cosmological parameters and a limited volume \citep[we use $\sigma_8=0.9$ in comparison to 0.8 
predicted by][]{Komatsu09}. 
Second, our models do not allow for a random scatter in the MR (when computing the CF), 
which should naturally exist at some level.
Third, assuming that the stellar mass of galaxies depends only on the mass of the host subhalo and halo is a 
significant simplification. From more complex models it is known that the large-scale environment might effect
stellar masses up to a few per cent \citep[the `assembly bias' effect][]{Gao05,Croton07}. Also, the time a galaxy
became a satellite might play a role \citep[][]{Yang12}. Obviously, an
accurate determination of the galaxy mass within a halo must take into account many properties of the merger history
of that halo. Since all these effects are predicted to be at a level of a few per cent, our approach of fitting the
data to a level of 0.1 dex might be secure enough.

In terms of modeling the WL signal, our approach includes an additional set of assumptions. We assume here that the lensing
signal is fixed only by the dark matter content. However, baryonic dynamics might change the dark-matter profiles
of subhaloes, and their location within a group. In addition, the mass of gas and stars can contribute directly
to the lensing signal \citep{Leauthaud12}. All these assumptions should be examined in the near future, as the 
observed data sets are getting more accurate, demanding more complex and general models.


\section*{Acknowledgments}

The Millennium Simulation databases used in this paper and the web application providing online 
access to them were constructed as part of the activities of the German Astrophysical Virtual Observatory.
We thank Gerard Lemson for his help in extracting the particle data from the Millennium simulation.
EN and SK acknowledge funding by the DFG via grant KH-254/2-1. We thank Rachel Mandelbaum for sharing her data 
in an electronic form and for enlightening discussions, Marcello Cacciato and Frank van den Bosch for helpful
discussions, and Bhaskar Agarwal for helpful comments on an earlier draft.
%

\bibliographystyle{mn2e}
\bibliography{ref_list}

\appendix
\section{Database of profile shapes}
\label{sec:app_prof}

We provide the values of the WL profiles, $\Delta\Sigma_c$ and $\Delta\Sigma_s$, averaged in bins of $\mh$ and $\mf$.
The original tables have a bin size of 0.1 dex in each variable (except for $\mh<10^{12}$ for which the bin size 
is 0.02 dex), so we only write a sub-sample of bins here. For example,
only half the values as a function of $r$ are written. In addition, the number of subhaloes summed over all 
the lines within a table does not accumulate to the number of subhaloes in the simulation.

\begin{table*}
\caption{The average WL profiles ($\Delta\Sigma_c$) for central subhaloes, as a function of $\mh$. Units of $\mh$ are
$\log\msun$, profile units are $h\msun$ pc$^{-2}$, and distances are given in $\log$ Mpc $h^{-1}$. 
$N$ is the number of objects used for deriving the average. }
\begin{center}
\begin{tabular}{lcccccccccccccc}
\hline $\mh$ & $N$ & $\log r=-1.5$ & -1.3 & -1.1 & -0.9 & -0.7 & -0.5 & -0.3 & -0.1 & 0.1 & 0.3 & 0.5 & 0.7 & 0.9  \\
\hline
11.0 & 179089 & 5.01 & 3.36 & 1.91 & 1.00 & 0.48 & 0.22 & 0.09 & 0.09 & 0.21 & 0.28 & 0.31 & 0.29 & 0.21 \\
11.2 & 112213 & 7.10 & 4.85 & 2.81 & 1.53 & 0.79 & 0.39 & 0.18 & 0.11 & 0.20 & 0.26 & 0.31 & 0.28 & 0.25 \\
11.4 & 77532 & 9.89 & 6.93 & 4.17 & 2.32 & 1.18 & 0.58 & 0.28 & 0.25 & 0.26 & 0.30 & 0.30 & 0.26 & 0.25 \\
11.6 & 48412 & 13.56 & 9.71 & 5.94 & 3.35 & 1.75 & 0.77 & 0.33 & 0.23 & 0.20 & 0.26 & 0.32 & 0.32 & 0.27 \\
11.8 & 33503 & 18.33 & 13.46 & 8.52 & 4.92 & 2.62 & 1.28 & 0.58 & 0.30 & 0.23 & 0.21 & 0.32 & 0.32 & 0.27 \\
12.0 & 63990 & 25.10 & 18.91 & 12.34 & 7.42 & 4.05 & 2.04 & 0.99 & 0.48 & 0.29 & 0.33 & 0.37 & 0.32 & 0.28 \\
12.2 & 73130 & 31.98 & 24.74 & 16.52 & 10.14 & 5.77 & 2.99 & 1.42 & 0.67 & 0.36 & 0.33 & 0.36 & 0.33 & 0.26 \\
12.4 & 48867 & 41.36 & 32.81 & 22.55 & 14.16 & 8.23 & 4.41 & 2.16 & 1.02 & 0.58 & 0.37 & 0.37 & 0.31 & 0.31 \\
12.6 & 32122 & 53.01 & 43.23 & 30.52 & 19.69 & 11.75 & 6.43 & 3.28 & 1.59 & 0.76 & 0.51 & 0.42 & 0.37 & 0.29 \\
12.8 & 21282 & 67.13 & 56.16 & 40.66 & 26.84 & 16.43 & 9.37 & 4.91 & 2.36 & 1.01 & 0.61 & 0.47 & 0.44 & 0.37 \\
13.0 & 13690 & 83.75 & 71.85 & 53.51 & 36.32 & 22.77 & 13.34 & 7.22 & 3.71 & 1.71 & 0.85 & 0.63 & 0.44 & 0.44 \\
13.2 & 8951 & 103.04 & 90.94 & 69.68 & 48.45 & 31.28 & 18.75 & 10.51 & 5.46 & 2.60 & 1.27 & 0.79 & 0.52 & 0.50 \\
13.4 & 5724 & 125.22 & 113.53 & 89.14 & 63.55 & 42.19 & 26.13 & 15.02 & 8.05 & 4.07 & 1.84 & 1.08 & 0.65 & 0.56 \\
13.6 & 3608 & 148.39 & 138.16 & 111.90 & 82.46 & 56.02 & 35.80 & 21.31 & 11.89 & 6.03 & 2.89 & 1.35 & 0.77 & 0.64 \\
13.8 & 2178 & 176.51 & 168.68 & 139.91 & 105.93 & 73.98 & 48.37 & 29.44 & 16.95 & 9.00 & 4.25 & 2.15 & 1.11 & 0.78 \\
14.0 & 1293 & 203.92 & 200.51 & 171.91 & 132.95 & 95.40 & 64.74 & 41.06 & 24.10 & 13.20 & 6.23 & 3.13 & 1.83 & 1.38 \\
14.2 & 736 & 232.74 & 234.49 & 205.29 & 163.84 & 121.11 & 84.05 & 55.00 & 34.06 & 18.88 & 9.79 & 4.37 & 2.34 & 1.28 \\
14.4 & 443 & 263.40 & 269.42 & 242.00 & 199.49 & 152.45 & 106.83 & 71.57 & 45.88 & 27.03 & 14.64 & 7.23 & 3.12 & 1.65 \\
14.6 & 225 & 299.72 & 311.95 & 288.70 & 246.01 & 196.56 & 142.86 & 97.80 & 62.76 & 38.82 & 21.34 & 10.78 & 5.40 & 2.45 \\
14.8 & 92 & 327.30 & 355.52 & 339.56 & 297.30 & 239.66 & 180.41 & 126.47 & 82.29 & 52.06 & 28.70 & 17.51 & 7.88 & 3.66 \\
15.0 & 31 & 382.92 & 405.43 & 383.33 & 330.55 & 282.55 & 224.75 & 162.40 & 114.04 & 76.38 & 44.70 & 22.77 & 12.08 & 6.02 \\
\hline
\end{tabular}
\end{center}
\label{tab:profiles_cent}
\end{table*}
\begin{table*}
\caption{The average WL profiles ($\Delta\Sigma_s$) for satellite subhaloes, as a function of $\mh$ and $\mf$. Here we use a model
with $\adf=3$, and location of unresolved subhaloes following their most bound particle. Mass units are
$\log\msun$, profile units are $h\msun$ pc$^{-2}$, and distances are given in $\log$ Mpc $h^{-1}$. 
$N$ is the number of objects used for deriving the average. }
\begin{center}
\begin{tabular}{lccccccccccccccc}
\hline $\mh$ & $\mf$ & $N$ & $\log r=-1.5$ & -1.3 & -1.1 & -0.9 & -0.7 & -0.5 & -0.3 & -0.1 & 0.1 & 0.3 & 0.5 & 0.7 & 0.9  \\
\hline
11.0 & 12.0 & 3833 & 7.24 & 7.31 & 6.50 & 4.79 & 3.18 & 1.99 & 0.95 & 0.39 & 0.14 & 0.26 & 0.46 & 0.39 & 0.28 \\
11.2 & 12.0 & 2604 & 9.62 & 9.18 & 7.36 & 5.02 & 3.10 & 1.76 & 1.02 & 0.72 & 0.27 & 0.22 & 0.43 & 0.26 & 0.34 \\
11.4 & 12.0 & 1547 & 12.13 & 10.58 & 7.79 & 5.25 & 3.50 & 1.85 & 0.84 & 0.36 & 0.22 & 0.48 & 0.61 & 0.53 & 0.38 \\
11.6 & 12.0 & 749 & 15.13 & 12.14 & 8.20 & 5.12 & 3.26 & 2.04 & 0.93 & 0.32 & 0.24 & 0.69 & 0.42 & 0.31 & 0.28 \\
11.8 & 12.0 & 394 & 17.82 & 14.17 & 9.52 & 5.81 & 2.99 & 1.91 & 1.18 & 0.32 & 0.03 & 0.32 & 0.19 & 0.59 & 0.26 \\
12.0 & 12.0 & 179 & 21.79 & 16.38 & 10.95 & 5.94 & 2.95 & 2.37 & 1.45 & 0.75 & 0.11 & 0.49 & 0.10 & 0.25 & 0.50 \\
11.0 & 13.0 & 4087 & 6.71 & 8.05 & 9.20 & 10.14 & 9.28 & 7.52 & 5.55 & 3.32 & 1.61 & 0.81 & 0.59 & 0.45 & 0.43 \\
11.2 & 13.0 & 3063 & 8.83 & 10.66 & 11.87 & 11.93 & 10.69 & 8.61 & 5.99 & 3.64 & 1.86 & 1.09 & 0.68 & 0.43 & 0.38 \\
11.4 & 13.0 & 1938 & 10.71 & 11.56 & 12.80 & 13.29 & 11.33 & 8.47 & 6.02 & 3.60 & 1.74 & 0.78 & 0.58 & 0.55 & 0.35 \\
11.6 & 13.0 & 1402 & 15.57 & 16.97 & 17.30 & 15.40 & 12.20 & 8.94 & 5.70 & 3.73 & 1.95 & 0.93 & 0.67 & 0.45 & 0.44 \\
11.8 & 13.0 & 849 & 21.61 & 22.62 & 20.89 & 18.11 & 13.50 & 8.82 & 5.35 & 3.44 & 1.51 & 1.09 & 0.79 & 0.63 & 0.51 \\
12.0 & 13.0 & 2443 & 33.06 & 32.37 & 27.67 & 21.35 & 14.90 & 9.79 & 6.19 & 3.67 & 1.86 & 1.07 & 0.79 & 0.48 & 0.48 \\
12.2 & 13.0 & 1641 & 43.87 & 39.53 & 31.16 & 22.69 & 15.11 & 9.88 & 5.97 & 3.76 & 1.86 & 0.95 & 0.53 & 0.53 & 0.33 \\
12.4 & 13.0 & 859 & 53.28 & 46.88 & 35.04 & 24.28 & 16.03 & 10.21 & 6.47 & 3.92 & 1.75 & 0.90 & 0.72 & 0.74 & 0.50 \\
12.6 & 13.0 & 441 & 59.65 & 51.41 & 38.08 & 26.50 & 16.76 & 9.73 & 6.71 & 3.97 & 2.32 & 1.09 & 0.18 & 0.72 & 0.65 \\
12.8 & 13.0 & 187 & 68.99 & 58.27 & 43.79 & 29.99 & 20.23 & 12.81 & 7.83 & 5.17 & 2.49 & 0.42 & 0.64 & 0.57 & 0.58 \\
13.0 & 13.0 & 20 & 75.13 & 68.09 & 50.83 & 32.43 & 19.01 & 11.15 & 7.99 & 4.45 & 2.34 & -0.44 & 1.30 & -0.17 & -0.43 \\
11.0 & 14.0 & 3408 & 5.71 & 7.79 & 10.99 & 14.55 & 17.62 & 18.24 & 17.64 & 14.85 & 10.67 & 6.22 & 3.20 & 1.88 & 1.39 \\
11.2 & 14.0 & 2646 & 7.31 & 8.60 & 11.71 & 15.68 & 18.92 & 19.13 & 17.35 & 15.02 & 10.79 & 6.08 & 3.01 & 1.83 & 1.33 \\
11.4 & 14.0 & 1544 & 9.36 & 11.58 & 15.09 & 20.25 & 22.43 & 21.66 & 19.38 & 15.25 & 11.04 & 6.10 & 3.22 & 1.75 & 1.40 \\
11.6 & 14.0 & 1140 & 11.18 & 13.01 & 16.13 & 18.96 & 20.92 & 21.19 & 19.20 & 15.12 & 10.50 & 6.04 & 3.04 & 1.90 & 1.18 \\
11.8 & 14.0 & 737 & 18.18 & 20.01 & 22.54 & 25.43 & 26.72 & 23.78 & 20.31 & 15.59 & 11.36 & 6.31 & 2.83 & 1.83 & 1.38 \\
12.0 & 14.0 & 2138 & 24.42 & 25.30 & 25.30 & 26.02 & 26.50 & 24.15 & 20.59 & 15.60 & 10.68 & 6.24 & 3.13 & 1.80 & 1.43 \\
12.2 & 14.0 & 1472 & 33.06 & 33.44 & 33.32 & 32.48 & 30.68 & 26.06 & 21.30 & 16.15 & 10.73 & 6.34 & 2.96 & 1.98 & 1.33 \\
12.4 & 14.0 & 987 & 43.81 & 42.95 & 40.61 & 37.59 & 33.99 & 29.34 & 22.91 & 16.76 & 11.57 & 6.15 & 3.45 & 2.00 & 1.32 \\
12.6 & 14.0 & 696 & 61.50 & 63.04 & 57.99 & 50.88 & 42.01 & 31.33 & 23.59 & 16.76 & 11.09 & 6.41 & 3.10 & 2.02 & 1.60 \\
12.8 & 14.0 & 439 & 77.28 & 78.17 & 73.24 & 62.17 & 49.66 & 38.85 & 26.85 & 17.81 & 11.22 & 5.56 & 3.63 & 2.05 & 1.48 \\
13.0 & 14.0 & 284 & 103.83 & 104.50 & 93.05 & 75.28 & 57.45 & 40.39 & 28.76 & 18.79 & 12.21 & 6.27 & 2.61 & 2.53 & 1.45 \\
13.2 & 14.0 & 218 & 138.47 & 135.86 & 115.26 & 89.87 & 64.76 & 46.35 & 31.48 & 20.34 & 12.07 & 5.93 & 3.12 & 2.48 & 1.42 \\
13.4 & 14.0 & 85 & 142.17 & 140.62 & 118.46 & 91.15 & 63.71 & 42.15 & 25.28 & 16.98 & 12.00 & 6.87 & 3.52 & 2.15 & 1.33 \\
13.6 & 14.0 & 59 & 174.08 & 168.09 & 137.78 & 101.39 & 70.55 & 46.91 & 31.81 & 17.64 & 10.97 & 7.43 & 4.33 & 1.94 & 1.52 \\
13.8 & 14.0 & 13 & 168.92 & 169.20 & 145.13 & 108.62 & 76.51 & 49.61 & 33.05 & 21.45 & 13.61 & 5.65 & 1.70 & 4.16 & 3.46 \\
14.0 & 14.0 & 1 & 225.79 & 210.90 & 183.90 & 147.66 & 98.33 & 64.17 & 47.14 & 27.89 & 9.89 & 1.68 & -3.17 & 6.38 & 6.88 \\
11.0 & 15.0 & 790 & 5.36 & 7.07 & 10.68 & 17.01 & 24.12 & 30.04 & 35.65 & 39.78 & 39.47 & 31.99 & 20.27 & 9.85 & 5.12 \\
11.2 & 15.0 & 602 & 6.51 & 8.37 & 11.23 & 17.63 & 24.81 & 33.02 & 39.34 & 40.35 & 38.39 & 31.55 & 20.65 & 10.36 & 5.15 \\
11.4 & 15.0 & 324 & 7.63 & 9.66 & 12.82 & 16.72 & 26.05 & 41.61 & 42.84 & 45.61 & 38.86 & 31.12 & 20.19 & 10.11 & 5.14 \\
11.6 & 15.0 & 250 & 12.95 & 14.19 & 17.54 & 22.39 & 35.76 & 40.53 & 43.01 & 41.36 & 38.80 & 31.28 & 20.30 & 10.37 & 5.01 \\
11.8 & 15.0 & 157 & 9.62 & 9.36 & 12.76 & 22.35 & 25.81 & 31.68 & 36.54 & 44.11 & 40.00 & 32.60 & 20.23 & 10.63 & 5.20 \\
12.0 & 15.0 & 449 & 26.57 & 26.79 & 27.06 & 33.02 & 40.83 & 45.99 & 47.99 & 44.18 & 40.66 & 31.24 & 19.56 & 9.93 & 5.15 \\
12.2 & 15.0 & 313 & 34.28 & 38.37 & 40.34 & 42.81 & 41.19 & 43.43 & 45.65 & 47.34 & 43.45 & 33.16 & 21.27 & 10.08 & 5.06 \\
12.4 & 15.0 & 218 & 43.08 & 37.16 & 31.10 & 34.07 & 41.06 & 43.35 & 43.18 & 42.31 & 41.64 & 33.71 & 21.06 & 10.61 & 5.24 \\
12.6 & 15.0 & 136 & 54.77 & 55.29 & 50.31 & 50.93 & 51.86 & 51.09 & 46.73 & 43.61 & 39.70 & 31.11 & 20.30 & 9.68 & 5.41 \\
12.8 & 15.0 & 98 & 55.28 & 59.64 & 57.88 & 56.11 & 49.22 & 56.06 & 59.90 & 53.27 & 47.88 & 34.52 & 20.74 & 10.68 & 5.33 \\
13.0 & 15.0 & 59 & 81.77 & 82.71 & 71.99 & 66.60 & 61.87 & 61.33 & 60.09 & 55.88 & 49.33 & 33.81 & 21.22 & 10.44 & 5.10 \\
13.2 & 15.0 & 37 & 106.61 & 102.69 & 94.42 & 82.73 & 76.34 & 66.32 & 58.70 & 51.29 & 41.92 & 32.29 & 19.89 & 9.44 & 4.74 \\
13.4 & 15.0 & 29 & 116.42 & 133.20 & 137.17 & 142.69 & 135.84 & 133.73 & 101.57 & 73.56 & 57.10 & 40.19 & 23.12 & 10.40 & 5.41 \\
13.6 & 15.0 & 10 & 204.44 & 249.74 & 235.13 & 213.34 & 181.80 & 145.15 & 116.86 & 88.08 & 58.36 & 37.27 & 18.72 & 9.72 & 5.26 \\
13.8 & 15.0 & 16 & 198.24 & 216.06 & 201.26 & 168.52 & 158.21 & 128.33 & 97.10 & 74.48 & 56.77 & 36.22 & 20.37 & 10.85 & 5.48 \\
14.0 & 15.0 & 7 & 277.20 & 288.71 & 273.05 & 251.70 & 204.72 & 171.46 & 121.79 & 85.65 & 59.05 & 41.00 & 22.40 & 9.29 & 5.53 \\
14.2 & 15.0 & 8 & 368.56 & 393.56 & 356.73 & 322.69 & 264.57 & 205.77 & 146.69 & 98.66 & 66.81 & 37.12 & 21.61 & 10.51 & 6.50 \\
14.4 & 15.0 & 2 & 254.02 & 382.42 & 490.39 & 436.54 & 357.63 & 277.31 & 189.11 & 116.40 & 73.52 & 35.53 & 21.29 & 8.93 & 5.02 \\
14.6 & 15.0 & 1 & 669.29 & 637.40 & 480.35 & 319.97 & 192.59 & 155.75 & 220.41 & 149.22 & 78.59 & 46.37 & 21.30 & 8.26 & 2.79 \\
\hline
\end{tabular}
\end{center}
\label{tab:profiles_sat}
\end{table*}

\label{lastpage}

\end{document}